\documentclass[acmsmall,screen]{acmart}

\usepackage{xspace}
\usepackage{soul}

\newcommand{\approach}{\textsc{TestPrune}\xspace}

\newcommand{\verified}{{SWE-Bench-Verified}\xspace}

\newcommand{\lite}{{SWE-Bench-Lite}\xspace}
\newcommand{\swebench}{{SWE-Bench}\xspace}

\newcommand{\added}[1]{#1}

\usepackage[ruled,vlined,linesnumbered]{algorithm2e}

\SetKwInput{KwInputs}{Inputs}                
\SetKwInput{KwOutputs}{Outputs}              
\SetKwInput{KwOutput}{Output}  

\usepackage[final,commentmarkup=uwave,commandnameprefix=always]{changes}
\usepackage{alltt}
\usepackage{url}
\usepackage{todonotes}
\usepackage{multirow}
\usepackage{cleveref}
\usepackage{wrapfig}
\usepackage{xspace}
\usepackage{paralist}
\usepackage{multirow}
\usepackage{subcaption}
\usepackage{tikz}
\usepackage{courier}
\usepackage{listings} 
\usepackage{array}
\usepackage{graphics}
\usepackage{lipsum}
\usepackage{wrapfig}
\usepackage{arydshln}
\usepackage{url}
\usepackage{alltt}
\usepackage{booktabs} 
\usepackage{xcolor}
\usepackage{pifont}

\newcommand{\cmark}{\textcolor{green!70!black}{\ding{51}}} 
\newcommand{\xmark}{\textcolor{red}{\ding{55}}} 

\definecolor{Gray}{gray}{0.3}
\tikzstyle{mybox} = [draw=black, very thick, rectangle, rounded corners, inner ysep=5pt, inner xsep=5pt, fill=gray!20]

\newcommand{\research}[2]{
    \smallskip
    \noindent
    \begin{tikzpicture}
        \node [mybox] (box){%
        \centering
        \begin{minipage}{.97\columnwidth}
        \fontsize{8.8}{10}\selectfont
        \textbf{RQ #1}. #2
        \end{minipage}
        };
    \end{tikzpicture}%
}

\newcommand{\findings}[2]{
    \smallskip
    \noindent
    \begin{tikzpicture}
        \node [mybox] (box){%
        \centering
        \begin{minipage}{.95\columnwidth}
        \fontsize{8.8}{10}\selectfont
        \textbf{Finding #1}. #2
        \end{minipage}
        };
    \end{tikzpicture}%
}

\AtBeginDocument{%
  }

\setcopyright{cc}
\setcctype{by}
\acmDOI{10.1145/3808148}
\acmYear{2026}
\acmJournal{PACMSE}
\acmVolume{3}
\acmNumber{FSE}
\acmArticle{FSE141}
\acmMonth{7}
\acmSubmissionID{fse26mainb-p613-p}
\received{2026-02-24}
\received[accepted]{2026-03-24}

\begin{document}

\title{Can Old Tests Do New Tricks for Resolving SWE Issues?}

\author{Yang Chen}
\authornote{Author was an intern at IBM at the time of this work.}
\email{yangc9@illinois.edu}
\orcid{}
\affiliation{%
  \institution{University of Illinois at Urbana-Champaign}
  \city{Champaign}
  \state{Illinois}
  \country{USA}
}

\author{Toufique Ahmed}
\email{tfahmed@ibm.com}
\orcid{}
\affiliation{%
  \institution{IBM}
  \city{Yorktown Heights}
  \state{New York}
  \country{USA}
}

\author{Reyhaneh Jabbarvand}
\email{reyhaneh@illinois.edu}
\orcid{}
\affiliation{%
  \institution{University of Illinois at Urbana-Champaign}
  \city{Champaign}
  \state{Illinois}
  \country{USA}
}

\author{Martin Hirzel}
\email{hirzel@us.ibm.com}
\orcid{}
\affiliation{%
  \institution{IBM}
  \city{Yorktown Heights}
  \state{New York}
  \country{USA}
}

\begin{abstract}



Test suites in real-world projects are often large and achieve high code coverage, yet they remain insufficient for detecting all bugs. The abundance of unresolved issues in open-source project trackers highlights this gap. While regression tests are typically designed to ensure past functionality is preserved in the new version, they can also serve a complementary purpose: debugging the current version. Specifically, regression tests can (1)~enhance the generation of reproduction tests for newly reported issues, and (2)~validate that patches do not regress existing functionality. We present \approach, a fully automated technique that leverages issue tracker reports and strategically reuses regression tests for both bug reproduction and patch validation. 

A key contribution of \approach is its ability to automatically minimize the regression suite to a small, highly relevant subset of tests. Due to the predominance of LLM-based debugging techniques, this minimization is essential as large test suites exceed context limits, introduce noise, and inflate inference costs. \approach can be plugged into any agentic bug repair pipeline and orthogonally improve overall performance. \added{As a proof of concept, we show that \approach leads to a $6.2\%-9.0\%$ relative increase in issue reproduction rate within the Otter framework and a $8.0\%-12.9\%$ relative increase in issue resolution rate within Agentless, SWE-Agent, and Trae agent} on SWE-Bench Lite and SWE-Bench Verified benchmarks. Compared to the benefits, the model API cost overhead of \approach is minimal, at \$0.02 and \$0.05 per SWE-Bench instance using GPT-4o and Claude-3.7-Sonnet models, respectively. 

\end{abstract}




\maketitle

\section{Introduction}
\label{sec:introduction}

The purpose of regression tests is to ensure the code that worked before keeps working: they check that correct existing behavior does not regress. New software engineering (SWE) issues typically reveal cases where existing behavior is wrong in ways not covered by regression tests. While this mismatch limits their effectiveness for detecting existing bugs, regression tests can still provide valuable signals for debugging, in particular, for guiding the reproduction of reported issues and validating candidate patches. However, exploiting regression suites in this way raises new challenges: they are often large, time-consuming to execute, and too noisy to be useful in full~\cite{yoo2012regression}. These challenges are exacerbated in the context of agentic SWE pipelines, where input length, inference cost, and reasoning accuracy all deteriorate with excessive or irrelevant tests~\cite{ahmed_et_al_2025,levy2024same}.

This paper proposes \approach, which automatically extracts a small, relevant subset of regression tests to make LLM-based debugging workflows more efficient and reliable. We formalize identification of relevant regression tests into the \emph{issue-based test-suite minimization} problem: given an issue description
and existing repository, \approach selects the minimum number of regression tests (to execute faster with lower likelihood of flakiness) that are relevant to the issue (likely to cover the eventual generated patches, especially newly added or modified statements). Unlike prior test-suite minimization techniques that select a subset of tests that maintain coverage or bug detection abilities of the existing tests~\cite{lin2018nemo}, \approach should use the relevance to the issue for guiding the optimization. This is, however, tricky, since the patch is \emph{unknown} at the time of test suite minimization. To overcome this challenge, \approach leverages a large language model~(LLM) to predict a list of suspicious methods given the issue description (Section~\ref{sec:suspicious-function}), and favors tests that cover suspicious methods during their execution during selection (Section~\ref{sec:greedy-algorithm}). 

The key idea of \approach is simple, yet not explored by prior techniques. There are, in fact, several agentic workflows that have used regression tests to assist with automated issue reproduction~\cite{ahmed_et_al_2025,wang_et_al_2025,applis_et_al_2026,cheng_et_al_2025} or resolution~\cite{xia_et_al_2025,ruan_zhang_roychoudhury_2025,wei_et_al_2025,jain_et_al_2025,li_et_al_2025,gao2025trae}. Without selecting a small relevant subset of regression tests, their performance is sub-optimal: Cheng et al.~\cite{cheng_et_al_2025} demonstrate that generating and executing large-scale bug reproduction tests~(BRTs) drastically increases inference time and dilutes the efficacy of downstream repair models—resulting in only a 28\% plausible reproduction rate, whereas more focused, filtered guidance improves performance significantly. Xia et al.~\cite{xia_et_al_2025} report that despite the agent generating the patch, the ranking of plausible patches based on existing regression tests was not accurate enough to rank the correct patch on top. 

Our comprehensive evaluation of \approach demonstrates it can minimize subsets of relevant tests to the issues of SWE-Bench Verified~\cite{chowdhury_et_al_2024} and SWE-Bench Lite~\cite{swebench-lite} problems, \added{showing improvements along two dimensions:  (1) \textbf{Runtime performance improvement:} \approach minimizes the size by over $1{,}000\times$ compared to the original full test suite, on average, with a precision of $0.63$ and coverage recall (the fraction of buggy lines covered by \approach regression tests relative to the lines covered by the whole test suite) of $0.71$. (2) \textbf{Efficacy in practice for reproduction test generation and patch validation:} providing tests selected by \approach to Otter~\cite{ahmed_et_al_2025}, an agentic reproduction test generation approach, compared to original regression tests, yielded a $6.2\%-9.0\%$ relative improvement in issue reproduction rate on SWE-Bench Lite and SWE-Bench Verified benchmarks. \approach tests also helped the patch selection step of Agentless~\cite{xia_et_al_2025} with a relative improvement in issue resolution rate of $9.4\%-12.9\%$. \added{For agentic systems, \approach increases the issue resolution rate of SWE-agent~\cite{jimenez_et_al_2024} and Trae-agent~\cite{gao2025trae} by $8.0\%$--$9.4\%$, while reducing the average cost per instance by $8.0\%$--$23.0\%$.} \textbf{Cost efficiency:} it only cost additionally \$0.02 and \$0.05 per SWE-Bench instance, using GPT-4o and Claude-3.7-Sonnet models, respectively.}

This paper makes the following notable contributions:
\vspace*{-0.8mm}
\begin{enumerate}
  \item Identifying the problem of issue-based test minimization, where the code that resolves the issue is not yet known, and formulating the problem's intrinsic success metrics.
  
  \item \approach, the first approach for localizing a minimized relevant set of regression tests existing in a code repository based on an issue description. Our artifacts are publicly available~\cite{artifact}.
  
  \item Conducting a comprehensive evaluation of the effectiveness and efficacy of \approach to improve automated issue reproduction and issue resolution.
\end{enumerate}
\vspace*{-0.8mm}
The paper title asks whether old tests can do new tricks. This paper demonstrates that the answer is yes, by showing how to select the right subset of old tests and how to leverage them for important downstream tasks. 
\section{Problem Statement and Challenges}
\label{subsec:problem}



We formally define the issue-based test minimization problem as follows:

\smallskip


\noindent \textbf{Given:}  
(1) A program $P=\{M_1, M_2, \ldots, M_p\}$ consisting of $p$ methods and each method $M_i=\{s_{1}^i, s_{2}^i, \ldots, s_{k_i}^{i}\}$ consisting of $k_i$ lines; 
(2) an issue description $d_\textrm{issue}$ and an unknown (latent) function $\mathcal{F}(d_\textrm{issue}) \subseteq P$ mapping the issue description to a subset of methods $M_is$;
and (3) a regression test suite $RT=\{t_1, t_2, \ldots, t_l\}$ with each test represented as a coverage vector $\vec{\mathit{cov}}_{t_j}=\langle c_{j_1}, c_{j_2}, \ldots, c_{j_p}\rangle$, such that $c_{j_i}$ is the set of lines in $M_i \in \mathcal{F}(d_\textrm{issue})$ covered by executing~$t_j$ (could be empty if $t_j$ does not cover~$M_i$).
The function $\mathcal{F}(d_\textrm{issue})$ is not directly observable. Instead, it represents the latent set of methods that will eventually be modified to resolve the issue. In practice, this set must be guessed from the issue description and original code base.

\smallskip


\noindent \textbf{Problem.}  Find a test suite subset
$\mathcal{R}\subseteq RT$ such that
for every other $\mathcal{T}\subseteq RT$,
\begin{enumerate}
  \item $\mathcal{R}$ covers at least as many lines in issue-relevant methods:
    $\displaystyle\Big|\bigcup_{t_j\in \mathcal{R}} \vec{cov}_{t_j}\Big|
    \ge \Big|\bigcup_{t_j\in\mathcal{T}} \vec{cov}_{t_j}\Big|$, and
  \item if $\mathcal{R}$ and $\mathcal{T}$ cover the same number of lines, then
    $\mathcal{R}$ is at least as minimal: $|\mathcal{R}| \le |\mathcal{T}|$.
\end{enumerate}

\smallskip

The goal of issue-based test-suite minimization is to find a minimal subset of regression tests that cover the code that will \emph{eventually be modified} to resolve the described issues. Therefore, a test case that covers more methods (or more lines of methods) that are potentially the culprit for the issues (i.e., those $\in \mathcal{F}(d_\textrm{issue})$) has a higher significance compared to other regression tests. The coverage term
in the problem definition allows us to characterize the importance of a test, i.e., assigning a higher significance to $M_i$s $\in \mathcal{F}(d_\textrm{issue})$, and among them, assigning a higher significance to those that cover more lines during execution. 

One of the challenges in this problem is that the issue descriptions have an associated latent set of contributing methods, which are not directly observed but can be predicted. That is, issue descriptions are mainly in natural language and may include embedded stack traces or code snippets to map to program methods. To overcome these challenges, \approach prompts LLMs to determine a list of suspicious methods to approximate $\mathcal{F}(d_\textrm{issue})$ (\S \ref{sec:suspicious-function}). 

Due to inherent intra- and inter-class dependencies, and hence, the prevalence of call chains in real-world projects, several tests in the regression test suite may cover the same method (or the same lines of a method). As a result, the regression test can be partitioned into subsets of $T_1$, $T_2$, $\ldots$, $T_x \subseteq RT$, such that any test $t_j$ belonging to $T_i$ covers the method $M_i \in \mathcal{F}(d_\textrm{issue})$. A representative set of tests
that covers all of the $M_i$s must contain at least one test from each $T_i$, and is called \emph{the hitting set} regression test partitions. Our issue-based test-suite minimization problem, therefore, is an instance of the \emph{weighted minimal hitting set} problem, which is NP-hard. 

Finding the optimal solution for an NP-hard problem may require exponential time or be infeasible. The goal of \approach is to select the minimal set of relevant sets to issue descriptions for use in agentic pipelines for reproduction test generation and patch validation, demanding a faster solution. For this reason, it adopts a greedy heuristic that sacrifices guaranteed optimality in exchange for significantly quicker computation (\S \ref{sec:greedy-algorithm}).

\approach provides the first dedicated approach to issue-based test minimization, leveraging large language models and coverage analysis to systematically identify a test set that is both minimal and relevant to the issue. Minimality is essential in our problem formulation to reduce execution cost and avoid test flakiness. Solving this problem is crucial for downstream tasks in automated software engineering. A minimized, issue-relevant regression suite enables:  
\begin{enumerate}
	\item \textbf{Reproduction Test Generation.} By providing focused guidance on existing tests that cover suspicious functions, it increases the likelihood that LLM-based systems can generate fail-to-pass reproduction tests.  
	\item \textbf{Patch Selection and Validation.} By ensuring that selected tests remain relevant to the issue, it improves the precision of patch validation, reducing both false positives and false negatives when ranking plausible patches, \added{while also serving as feedback when incorporated into a patch refinement loop}.  
\end{enumerate}

\section{Approach}
\label{sec:approach}
 \begin{figure}[htbp]
  \centering
  \includegraphics[width=0.9\linewidth]{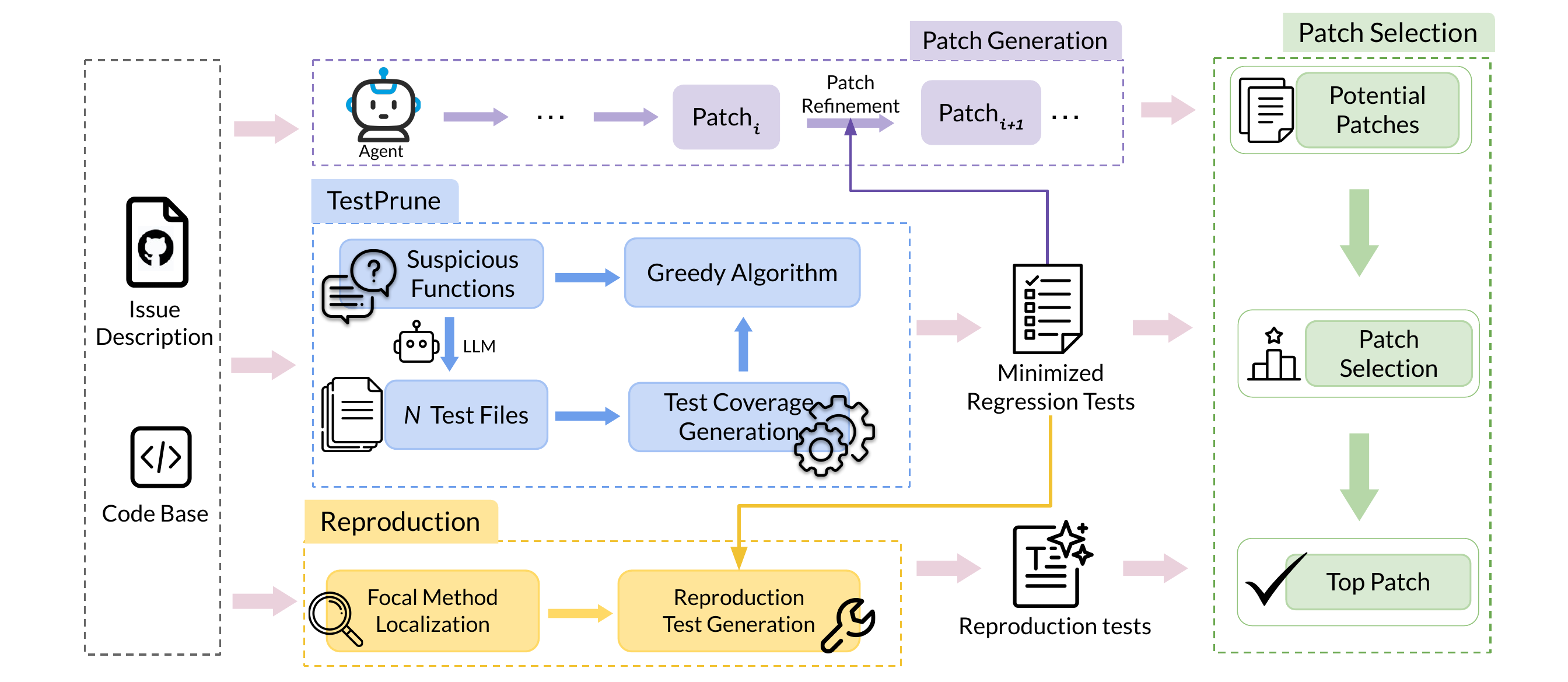}
  \vspace{-10pt}
  \caption{Overview of \approach and clients, reproduction test generation (yellow), patch generation (purple), and patch selection (green), in the end-to-end pipeline of fixing open-source issues \added{(added patch generation components)}}
  \vspace{-10pt}
  \label{fig:approach}
\end{figure}

\noindent
Figure~\ref{fig:approach} shows the workflow of \approach and its downstream applications. Starting from a GitHub issue and the corresponding code base, \approach first prompts an LLM to retrieve a set of suspicious functions and potentially relevant test files. It then combines coverage information with a greedy algorithm to select a minimized subset of regression tests. \added{These minimized regression tests can be applied to one or more downstream clients: any reproduction test generation workflows (e.g., Otter~\cite{ahmed_et_al_2025}), agentic patch generation (e.g., SWE-agent~\cite{jimenez_et_al_2024} or Trae Agent~\cite{gao2025trae}), and non-agentic patch generation and selection workflows~(e.g., Agentless~\cite{xia_et_al_2025}).}

\subsection{\approach}
\subsubsection{\textbf{Suspicious Function Localization.}}
\label{sec:suspicious-function}
\begin{wrapfigure}{r}{0.45\columnwidth}
    \vspace{-20pt}
\includegraphics[width=0.45\columnwidth]{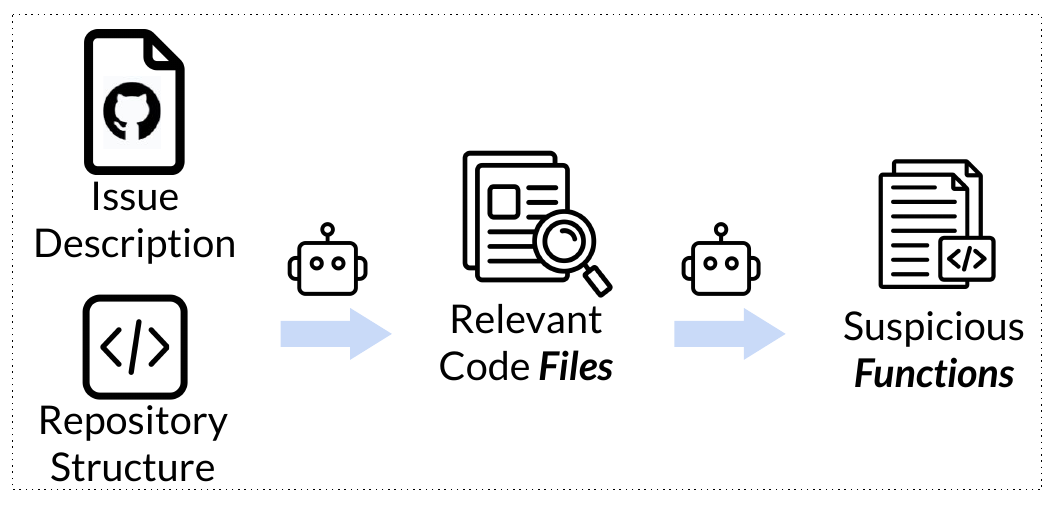}
\vspace{-20pt}
    \caption{Suspicious function localization}
    \vspace{-10pt}
    \label{fig:suspicious-function}
\end{wrapfigure}

Figure~\ref{fig:suspicious-function} illustrates the two-step localization procedure in \approach. Given a GitHub issue description and the structure of the code repository, it first localizes the issue to potentially relevant code files. Next, it prompts the model to narrow down the search to suspicious functions within those files.

In contrast to existing agentic program repair techniques~\cite{xia_et_al_2025,zhang2024autocoderover,gao2025trae, li_et_al_2025}, 
which employ a multi-step localization strategy, \approach focuses on identifying tests that are most relevant to the issue. As a result, it does not attempt to further localize edits at the line level. The rationale is that fixing the same issue does not always require modifying the exact same lines of code. Overly restricting localization to line-level edits risks overlooking tests that capture the right functionality.

\begin{wrapfigure}{r}{0.43\columnwidth}
    \vspace{-10pt}
\includegraphics[width=0.4\columnwidth]{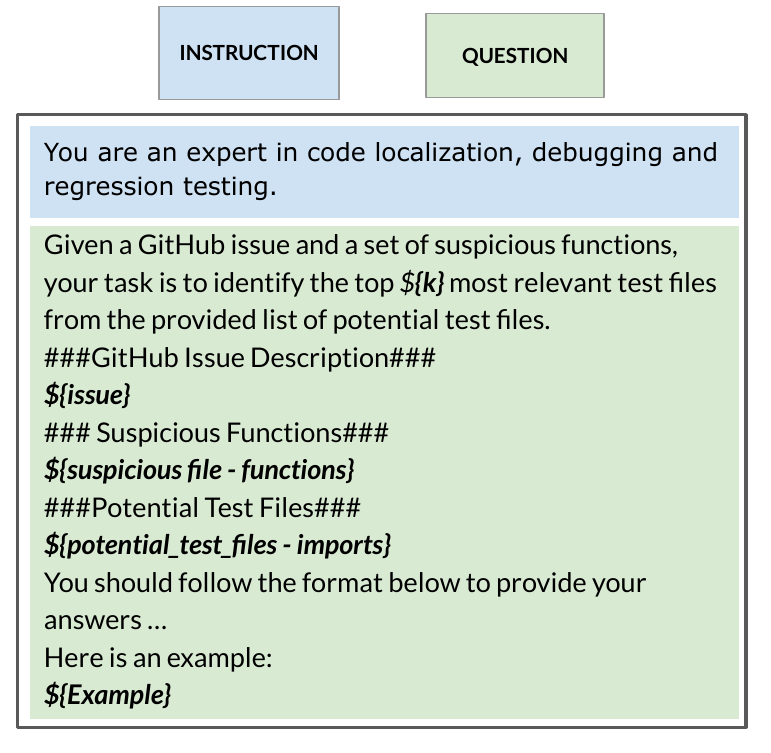}
\vspace{-15pt}
    \caption{Prompt template of test file selection}
    \label{fig:prompt-testfile-selection}
    \vspace{-20pt}
\end{wrapfigure}

\subsubsection{\textbf{Test File Retrieval and Coverage Generation.}}
\label{sec:test-file}
After identifying suspicious functions, \approach proceeds to locate tests related to those functions. 
Since test suites in open-source repositories can contain thousands of tests (see Section~\ref{sec:rq1}), exhaustively generating and mapping line-level coverage for all of them is expensive. To address this, \approach first prompts the LLM to retrieve the top $k$ candidate test files. Figure~\ref{fig:prompt-testfile-selection} shows the template used for this step. Specifically, it provides the model with (1) the GitHub issue description, (2) the suspicious functions together with their file paths, and (3) the list of test files in the repository along with the dependencies they import. Including the imported modules in the prompt provides the LLM with additional semantic context about the dependencies exercised by each test file, thereby helping it reason about which test files are more relevant. 
Once the candidate test files are retrieved, \approach collects all the passing test methods from these files and generates line-level coverage mappings to the suspicious functions. If no coverage is obtained, i.e., none of the tests exercise the suspicious functions, \approach reprompts the model for additional test files and repeats until non-empty coverage is collected.

\subsubsection{\textbf{Greedy Algorithm.}}
\label{sec:greedy-algorithm}
After collecting line-level coverage information of all suspicious functions, a greedy algorithm further minimizes the test set. \approach implements two greedy algorithms for test minimization: greedy-additional and greedy-total. \added{They differ in their optimization objectives. Greedy-additional recomputes coverage at each iteration using only the uncovered lines. In contrast, Greedy-total evaluates each candidate based on its overall coverage across all lines, regardless of whether previously selected tests already cover those lines.}

\begin{wrapfigure}{r}{0.62\textwidth}
\vspace{-13pt}
\small
\begin{algorithm}[H]
\caption{Greedy-\textit{Additional} Algorithm of Regression Tests Minimization}
\label{alg:greedy-add}
\SetAlgoLined
\SetKwInOut{Input}{Input}
\SetKwInOut{Output}{Output}

\Input{Suspicious functions $\mathcal{F}$; candidate tests $\mathcal{T}$; 
}
\Output{Minimized regression tests $\mathcal{R}$;}

$\mathcal{R} \leftarrow \emptyset$\;
$\mathcal{L} \leftarrow \textsc{AllLines}(\mathcal{F})$; 

\While{$\mathcal{L} \neq \emptyset$}{
  \ForEach{$t \in \mathcal{T}$}{
    $coverage(t) \leftarrow |\textsc{getCoverage}(t) \cap \mathcal{L}|$; 
  }
  $tests^\star \leftarrow \max_{t \in \mathcal{T}} coverage(t)$\;
    $\mathcal{T} \leftarrow \mathcal{T} \setminus \{tests^\star\}$\;
  $\mathcal{L} \leftarrow \mathcal{L} \setminus \textsc{getCoverage}(tests^\star)$; 
  
\eIf{len$(tests^\star) \leq 3$}{
  $\mathcal{R} \leftarrow \mathcal{R} \cup \{tests^\star\}$\;
}{
  $\mathcal{R} \leftarrow \mathcal{R} \cup \textsc{ModelBreakTie}(tests^\star)$\;
}

}
\Return $\mathcal{R}$\;
\end{algorithm}
\vspace*{-14pt}
\end{wrapfigure}

\vspace{10pt}
\noindent\textbf{Greedy-Additional Algorithm.}
Algorithm~\ref{alg:greedy-add} outlines this process. It takes the set of suspicious functions and candidate test methods as input, and produces a minimized regression suite as output. The algorithm begins by initializing the uncovered line set with all lines from the suspicious functions (line~2). In each iteration, it computes the coverage of each candidate test over these lines (lines $4$--$5$) and selects those achieving the highest coverage (line $7$). The selected tests are then removed from the candidate pool, and their covered lines are eliminated from the uncovered set (lines~$8$--$9$). When multiple tests achieve the same maximum coverage, if fewer than three tests exist in the tie, they are directly added to the minimized suite (lines~$10$--$11$); otherwise, an LLM is invoked to break the tie and select top candidates among them (lines~$12$--$13$). \added{We select more than one test because choosing a single test at random may be insufficient: different tests can exercise distinct inputs, assertions, or API usages even when they achieve the same coverage. Selecting multiple complementary tests improves both reproduction-test generation and patch validation by providing broader test behaviors.} This process continues until all lines are covered or no further useful tests remain. In the end, the minimized regression tests are returned.

\begin{wrapfigure}{r}{0.6\textwidth}
\vspace{-10pt}
\small
\begin{algorithm}[H]
\caption{Greedy-\textit{Total} Algorithm for Regression Test Minimization}
\label{alg:greedy-total}
\SetAlgoLined
\SetKwInOut{Input}{Input}
\SetKwInOut{Output}{Output}

\Input{Suspicious functions $\mathcal{F}$; candidate tests $\mathcal{T}$;}
\Output{Minimized regression tests $\mathcal{R}$;}

$\mathcal{R} \leftarrow \emptyset$\;
$\mathcal{C} \leftarrow \emptyset$\; 
\ForEach{$test \in \mathcal{T}$}{
  $\textit{Cov}(test) \leftarrow \bigl|\textsc{getCoverage}(test)\bigr|$; 
}
$\textit{noImprove} \leftarrow 0$\;

\While{$\mathcal{T} \neq \emptyset$ \textbf{and} $\textit{noImprove} < 3$}{

$\mathcal{C}_{\text{prev}} \leftarrow \mathcal{C}$\;

  $tests^\star \leftarrow \max_{test \in \mathcal{T}} \textit{Cov}(test)$\;
  

$\mathcal{C} \leftarrow \mathcal{C}_{\text{prev}} \cup \textsc{getLinesCoveredBy}(tests^\star)$\;

  \uIf{$|\mathcal{C}| > |\mathcal{C}_{\text{prev}}|$}{
    $\textit{noImprove} \leftarrow 0$\;
      \eIf{len$(tests^\star) \leq 3$}{
  $\mathcal{R} \leftarrow \mathcal{R} \cup \{tests^\star\}$\;
}{
  $\mathcal{R} \leftarrow \mathcal{R} \cup \textsc{ModelBreakTie}(tests^\star)$\;
}
  }
  \Else{
    $\textit{noImprove} \leftarrow \textit{noImprove} + 1$\;
  }

    $\mathcal{T} \leftarrow \mathcal{T} \setminus \{tests^\star\}$\;

}
\Return $\mathcal{R}$\;
\end{algorithm}
\vspace{-10pt}
\end{wrapfigure}
\noindent\textbf{Greedy-Total Algorithm.} 
We further introduce a greedy-\textit{total} variant algorithm (Algorithm~\ref{alg:greedy-total}). 
Unlike the greedy-\textit{additional} algorithm (Algorithm~\ref{alg:greedy-add}), it does not consider test coverage against the uncovered lines in previous iterations; instead, it ranks tests by their \emph{total} coverage of all lines in suspicious functions. 
The algorithm first collects each test’s total number of covered lines (lines~3--5). At each iteration, it  identifies the set of tests tied for the highest total coverage (line~9), and updates the cumulative coverage with the lines covered by this set (line~10).

If coverage increases, the non-improvement counter is reset (lines~11--12) and tie handling proceeds: when the tie size is at most three, those tests are added directly to the minimized suite (lines~13--14); otherwise, an LLM breaks the tie and selects the top candidate test methods (lines~15--16). 
If no new coverage is added, the non-improvement counter is incremented (lines~18--19). 
The tests in tie are removed from the pool (line~21) at the end of each iteration.
This process terminates after three consecutive non-improving iterations or when no test methods remain, and finally returns the minimized regression test suite (line~23).

\begin{figure}[t]
    \centering
    \includegraphics[width=0.99\columnwidth]{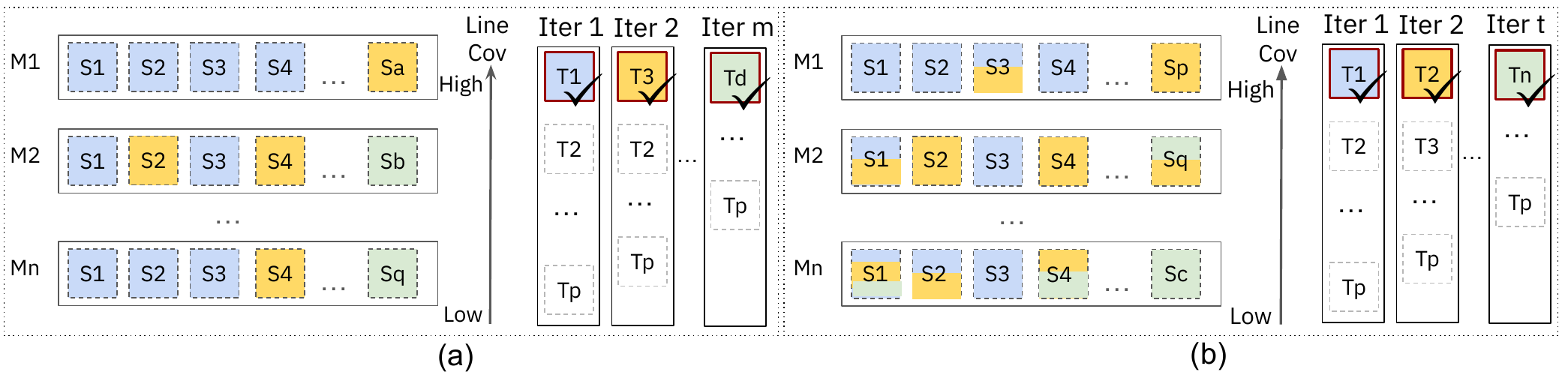}
    \vspace{-8pt}
    \caption{Example of greedy-additional algorithm (a) and greedy-total algorithm (b). $M$ denote suspicious functions,  $S$ are their lines, and $T$ denotes tests}
    \vspace{-15pt}
    \label{fig:greedy-example}
\end{figure}

Figure~\ref{fig:greedy-example} highlights the difference between the two greedy algorithms. 
For greedy-\textit{additional}, after \emph{Iter~1} selects $T_1$, the lines covered by $T_1$ are removed from the uncovered set; the coverage of the remaining tests is recomputed on the remaining lines, which can reshuffle the ranking (e.g., $T_3$ becomes the best choice in \emph{Iter~2}). This repeats until no test adds new coverage. 
For greedy-\textit{total}, tests are ranked once by their \emph{total} coverage over all suspicious lines and then taken in that fixed order across iterations (e.g., $T_1$, then $T_2$). During each iteration, previously covered lines are \emph{not} removed. The process stops after three consecutive non-improving iterations or when no more tests remain in the loop.


\vspace{-10pt}
\subsection{Reproduction Test}
\added{As shown in \Cref{fig:approach}, one downstream application of \approach is reproduction test generation, which in turn is essential for patch selection.}
Existing techniques either use zero-shot prompting of LLMs for reproduction test generation or use existing tests as additional context for reproduction test generation. Recent studies~\cite{mundler_et_al_2024, ahmed_et_al_2025} have shown that zero-shot approaches perform worse than methods that use relevant parts of the existing codebase as context. 
More advanced reproduction test generations still suffer from incompleteness, i.e., missing relevant functions to test due to the initial name-based file or function localization step~\cite{ahmed_et_al_2025}. \approach can mitigate the limitations of prior reproduction test generation work in two complementary ways: (1)~providing the minimized regression tests to give a richer context to the LLMs/agents of the reproduction workflow; and (2) in agentic systems, e.g., Otter~\cite{ahmed_et_al_2025}, presenting the agent with both focal localization and test localization during the self-reflective action planning phase (see~\Cref{background:otter}). We refer to this new variant a ``TestPrune''. If any regression test is already present in the prior localization, \approach drops it from the context to avoid duplication and reduce the false-positive rate (\S \ref{result:rq3}).




\subsection{Patch Selection and Validation}
\added{As shown in \Cref{fig:approach}, two other downstream applications of \approach are patch refinement and patch selection during and after generating a patch for program repair.
Existing program repair techniques can be categorized into two groups: (1)~Agentic systems, which are autonomous agents that iteratively plan, invoke tools, and refine their actions based on execution feedback to bug localization and patch generation~\cite{jimenez_et_al_2024, gao2025trae}. During the process, test execution results serve as critical information to guide subsequent decisions and actions.  These systems typically submit a single final patch in the end. (2)~Non-agentic systems~\cite{xia_et_al_2025}, which employ a multi-stage patch selection and validation.} At the first step, they run all or selected regression tests on all generated patches and only keep those with the lowest number of regression test failures. At the next run, if applicable, they execute reproduction tests on these patches, selecting any patch that turns failing tests into passing ones. If no patch qualifies, they fall back on the regression test results for ranking the patches. In case of a tie, they either randomly select a patch or follow a majority-voting strategy to choose the final patch. 
\approach can enhance patch selection and validation faster and more effectively: by minimizing to more relevant regression tests, \added{\approach focuses on running tests that are highly likely to be related to the code, thereby improving patch generation for agentic systems by providing more related test execution feedback, and enhancing patch selection for non-agentic systems by excluding patches that break correct functionality and reducing the noise from unrelated tests that could otherwise favor incorrect patches.}


\section{Evaluation}
\label{sec:evaluation}

\subsection{Research Questions}
\label{sec:reserach-question}

This work proposes a dedicated solution to the issue-based test minimization problem. In addition, it aims to investigate how minimized test sets aid in issue reproduction and issue resolution. As discussed in the problem statement in~\Cref{subsec:problem}, the test set should be minimal to ensure efficiency, and it should also be relevant.
The first research question explores the level of minimization we can achieve and how it impacts the overall execution time.

\research{1}{To what extent can \approach reduce the number of tests to run and how efficient are the selected regression tests with respect to execution time?}

The second research question investigates the relevance of our tests to the issue description. Note that a function can be covered by multiple tests, and some tests may not even touch the line(s) that need to be covered. We aim to find out how capable our selected tests are at covering the suspicious function to expose regressions in the newly proposed candidate patches. We use two metrics, precision and coverage recall, to investigate this research question (detailed in~\Cref{subsec:rq2}).

\research{2}{To what extent are the minimized regression tests relevant to the buggy functions with respect to precision and coverage recall?}

As the title suggests, we want to see how old tests can perform new tricks in downstream tasks such as issue reproduction and issue resolution. This paper shows ways to incorporate our minimized set into issue reproduction, and we observe how it can improve issue reproduction capability. We use the fail-to-pass~(\( F\!\!\to\!\!P \)) rate to evaluate this research question.

\research{3}{How effectively can \approach assist in generating reproduction tests?}

Finally, we show how our selected tests help in patch validation to improve the performance of existing issue resolution systems.
\added{\approach supports issue resolution for patch validation across agentic and non-agentic approaches. We evaluate both categories of techniques: (1) Agentic systems~\cite{jimenez_et_al_2024, gao2025trae}, where an agent loop autonomously generates actions towards patch generation; we integrate \approach-selected regression tests to give validation feedback during that loop. (2)~Non-agentic systems, such as Agentless~\cite{xia_et_al_2025}, that often produce tens of candidate patches for a single issue; we leverage \approach-selected regression tests to help pick a single best patch. Following prior work, we adopt \emph{issue resolution rate} as the primary effectiveness metric.
}


\research{4}{How effectively can \approach aid in patch selection and validation during issue resolution?}

\subsection{Experiment Setup.}
To avoid long test suite runs while ensuring the quality of selected tests, we set $k=10$ (Section~\ref{sec:test-file}) in \approach after performing small-scale empirical experiments to balance efficiency and test quality. We conducted experiments on two datasets: \lite~\cite{swebench-lite} and \verified~\cite{swebench-verified}, both derived from \swebench~\cite{jimenez_et_al_2024}. We used two models, GPT-4o~\cite{hurst2024gpt} and Claude-3.7-Sonnet~\cite{claude}, with temperature set to $0.8$ following best practice in prior work~\cite{xia_et_al_2025}. For reproduction test generation, we used Otter~\cite{ahmed_et_al_2025} \added{, which corresponds to the \textit{Reproduction} component in Figure~\ref{fig:approach}}. 
\added{For agentic patch generation, we integrate \approach into two top-performing open-source agents on \lite and \verified with their original underlying models.  
Specifically, we use SWE-agent~\cite{jimenez_et_al_2024} with Claude-Sonnet-4 on \lite and Trae-agent~\cite{gao2025trae} with Claude-Sonnet-3.7 on \verified; these correspond to the \textit{patch generation} agents at the top of Figure~\ref{fig:approach}.
Since Trae-agent leverages test-time scaling in its approach, which substantially increases execution cost, we disable test-time scaling, report and compare results from a single run, following a similar evaluation in related work~\cite{xia2025livesweagent}.} 
For patch selection, we directly leveraged the buggy function localization and patches provided by Agentless~\cite{xia_et_al_2025}, \added{which corresponds to the \textit{patch selection} component in Figure~\ref{fig:approach}}. We briefly discuss these two approaches and highlight their limitation for better analysis of the results. 

\subsubsection{Agentic Patch Generation Workflows: SWE-agent and Trae Agent}
\label{background:agents}
\added{We select SWE-agent and Trae Agent as they are open-source agents ranked within the top-5 on the \lite and \verified leaderboards, respectively. We briefly describe each agent and discuss how we integrate \approach into their workflows.}

\added{SWE-agent~\cite{jimenez_et_al_2024} is an agent-based framework that equips an LLM with a custom Agent-Computer Interface (ACI) to autonomously resolve GitHub issues. Given an issue description, SWE-agent iteratively navigates the repository, searches for relevant code, edits files, and runs tests through a sequence of tool-invoked actions. The agent operates within a sandboxed environment and follows a ReAct-style~\cite{yao2023react} loop of reasoning, action, and observation. SWE-agent relies on the agent itself to decide which tests to run and when, without a dedicated regression testing mechanism. As a result, patches may break existing functionality if the agent fails to identify and execute the appropriate tests during its exploration.}

\added{Trae Agent~\cite{gao2025trae} further improves performance through test-time scaling, where multiple candidate solutions are generated, and the best is selected via voting. Specifically, it employs a tester agent and a selector agent to rank patches among multiple candidates, with the tester agent filtering regression tests for this process. However, the regression tests are limited to the patch ranking stage during test-time scaling and are not applied during patch generation.}

\begin{figure}[t]
    \centering
    \includegraphics[width=0.92\columnwidth]{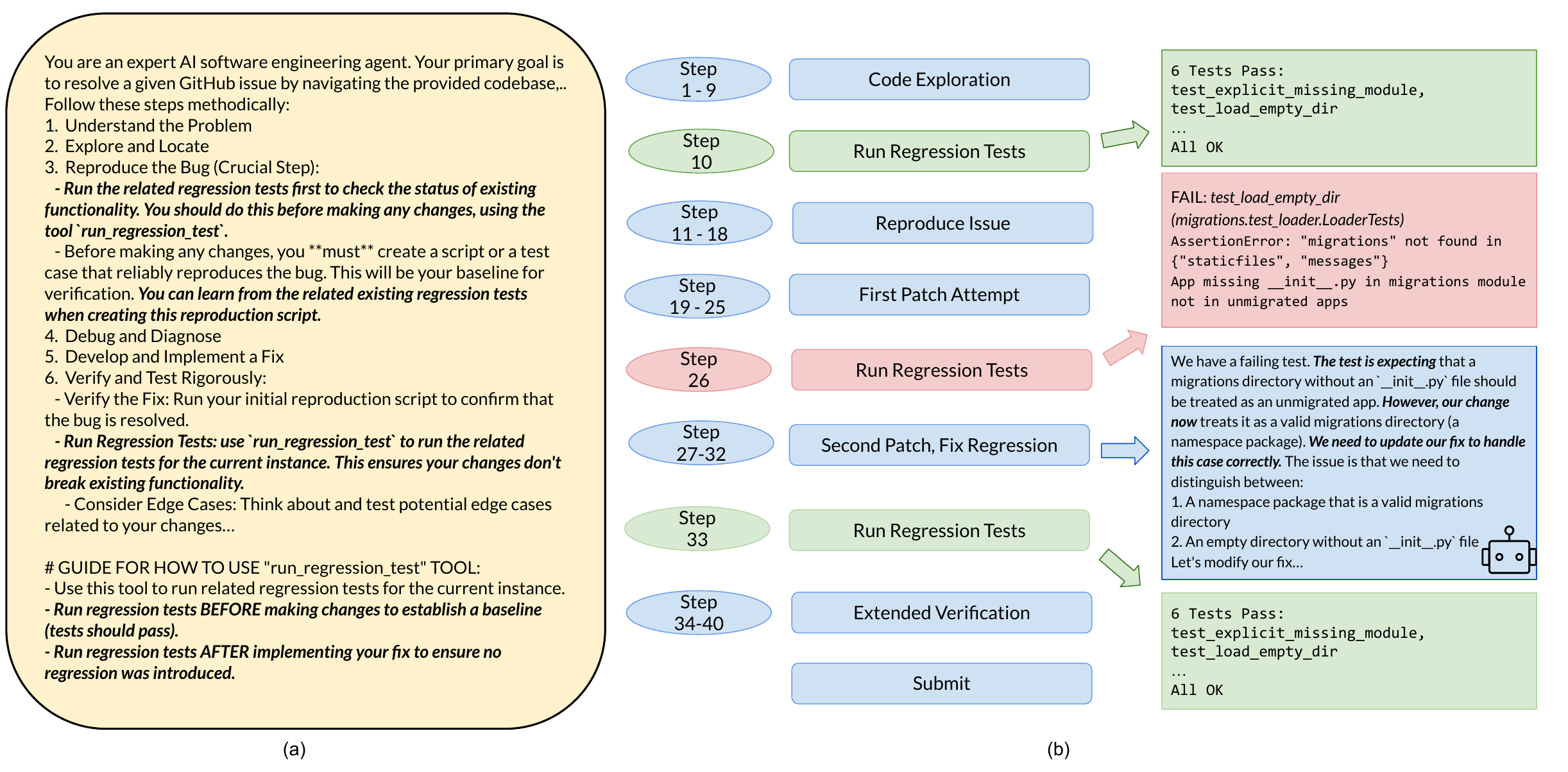}
    \vspace{-10pt}
    \caption{Agentic workflow with regression tests: (a) modified system prompt, and (2) example newly resolved instance django\_\_django-11141 by Trae Agent}
    \label{fig:test-design}
    \vspace{-10pt}
\end{figure}

\added{To integrate \approach-selected minimized regression tests into agents, we implement a tool \texttt{\small run\_regression\_tests} that collects the relevant regression tests for a given instance and executes them. We provide a detailed description of the tool's functionality in the system prompt, instructing the agent to run regression tests both at the beginning of issue investigation and after generating a candidate patch. The minimized regression tests can also be referenced when constructing reproduction tests.
Figure~\ref{fig:test-design}~(a) highlights the key modifications to the system prompt related to regression testing in Trae Agent (we apply similar changes to SWE-agent). Figure~\ref{fig:test-design}~(b) presents an example instance that was previously unresolved but is successfully fixed with the assistance of regression testing.
In this example, after the initial exploration phase, the agent runs the relevant regression tests and all pass (Step~10). However, after the first patch is generated (Step~25), one regression test \texttt{\small test\_load\_empty\_dir} fails (Step~26). Based on the failure message, the agent reflects on the discrepancy between expected and actual behavior (gracefully handling an empty directory), introduces additional changes to prevent the regression (Steps 27-32), and ultimately produces a correct patch, demonstrating how regression test feedback during patch generation can guide the agent toward a fix that resolves the issue.}

\subsubsection{Patch Selection and Validation Workflow: Agentless}
\label{background:agentless}

We chose Agentless due to the availability of its artifacts. 
Agentless employs a hierarchical bug localization process. Initially, it identifies suspicious files using both LLM-based prompting and embedding-based retrieval (with filtering to exclude irrelevant files). Next, it refines the search to relevant classes or functions using a concise skeleton representation of the files. Finally, it pinpoints precise edit locations with line-level accuracy.
For repair, Agentless generates multiple patches in a lightweight Search/Replace diff format rather than rewriting entire code blocks. This approach reduces hallucinations and improves cost-efficiency. Agentless generates 40 candidate patches for each issue, making patch validation a critical task, which the authors address using both reproduction and regression tests.

Agentless automatically generates reproduction tests and combines them with regression tests to select the best patch through majority voting. For regression testing, Agentless first runs all existing tests in the repository to identify a set of passing tests in the original codebase. It provides the list of passing tests to the LLM and asks it to identify any tests that should not be used to verify whether the issue has been correctly fixed. After removing the LLM-identified tests, Agentless obtains a final subset of old regression tests. Then it runs them on all generated patches, retaining only those with the fewest regression failures.

Running the whole test suite is time-consuming and can take more than half an hour for each issue, making it harder to apply in real development settings. Tests can fail due to flakiness~\cite{wang2025solved}  and dependency issues. Failing tests that are irrelevant to the issue can mislead the patch validation process. Besides, repeating test execution to mitigate the impact of test flakiness is not desirable due to time and resource constraints.
To address this problem, we propose replacing the Agentless test selection step with \approach. Running only the selected test subset from \approach saves time by eliminating the need to run the entire test suite, and provides signals from tests that are more relevant to the issue description.
Our results show that, apart from saving time, this small subset also increases patch validation accuracy in Agentless.

\subsubsection{Reproduction Test Generation Workflow: Otter}
\label{background:otter}

Besides issue resolution, this paper explores issue reproduction as a second downstream client for test minimization.
This is also an active research field with several approaches~\cite{ahmed_et_al_2025,wang_et_al_2025,cheng_et_al_2025}
leveraging old regression tests while generating new reproduction tests.
The metric \mbox{$F\!\!\to\!\!P$} measures the rate of generated reproduction
tests that are fail-to-pass, meaning they fail on the original code to
reproduce the issue and pass on the new code to validate the fix.
We use Otter~\cite{ahmed_et_al_2025} as a representative issue reproduction
system because of its strong \mbox{$F\!\!\to\!\!P$} rate.

Otter is a pipeline that combines LLM-based and rule-based steps. It consists of three main components: a localizer, a self-reflective action planner, and a test generator. The localizer identifies relevant files, test functions, and focal functions by querying the LLM with context-aware prompts. The action planner then iteratively constructs a plan of actions—categorized as read, write, and modify—guided by a self-reflective loop and validation checks. Finally, the test generator creates complete test functions using localized code and structural cues such as test signatures and imports.
It also includes mechanisms for repairing hallucinated or incomplete imports and integrates static analysis tools (e.g., the Flake8 Python linter) to ensure syntactic and semantic correctness. 

Otter++ enhances Otter by generating five test candidates using heterogeneous prompts, each varying in the inclusion of localized context. It selects the best candidate by analyzing execution outcomes (e.g., assertion failures vs.\ syntax errors) on the original code. Otter++ achieves higher \mbox{$F\!\!\to\!\!P$} rates than prior systems, while remaining cost-efficient and generalizable. The five prompt variants in Otter++ are planner, full, testLoc, patchLoc, and none. For example, the ``full'' prompt variant presents the model with both localizations~(focal and test) and asks it to generate the tests, while the ``testLoc'' variant presents the model with test localization but does not expose the focal localization.

One of the key limitations in Otter is that localization depends solely on file names and function names, not on real execution. Using \approach, we can expose more functions relevant to the issue to the planner, which can help create a better plan in the planning phase. In this work, we found that incorporating \approach can improve Otter's performance (detailed results in~\Cref{result:rq3}).

\subsubsection{Evaluation Benchmarks: SWE-Bench Verified and SWE-Bench Lite}

SWE-Bench~\cite{jimenez_et_al_2024} is a benchmark that tests how well LLMs can fix real GitHub issues by looking at an issue description and the old version of the code, then creating a patch that solves the problem. To make this easier and faster to use, SWE-Bench Lite provides a smaller set of 300 carefully chosen tasks that still cover the same variety and difficulty as the full benchmark, but with lower cost and quicker turnaround for researchers. LLM agents have made steady progress, but some tasks in the original benchmark are too hard or even impossible, which can make models look weaker than they really are. To address this, OpenAI worked with the creators of SWE-Bench on a new release with $500$ samples (SWE-Bench Verified~\cite{chowdhury_et_al_2024}) that gives more reliable evaluations, helping researchers better measure the ability of models for solving SWE tasks autonomously.

\begin{wrapfigure}{r}{0.3\columnwidth}
\includegraphics[width=0.3\columnwidth]{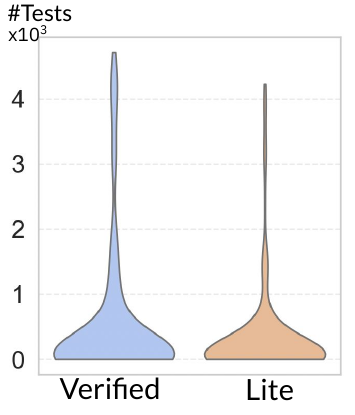}
\vspace{-20pt}
    \caption{Statistics of golden tests}
    \vspace{-5pt}
    \label{fig:golden-stat}
\end{wrapfigure}

We used SWE-Bench Lite and SWE-Bench Verified for both issue reproduction and issue resolution. For issue reproduction evaluation, there are two similar benchmarks: SWT-Bench~\cite{mundler_et_al_2024} and TDD-bench Verified~\cite{ahmed_et_al_2024-tdd}. 
Both benchmarks indicate whether a generated test reproduces the issue by checking whether the test fails on the existing code base ($c_\mathrm{old}$) before applying the golden patch and passes afterward. One difference between their implementations is that TDD-Bench only
runs contributing tests (tests that were added or updated), whereas SWT-bench runs the whole test files that contain the contributing tests. 
We use the harness of TDD-bench Verified because of its speed, but instead of evaluating on the 449 samples proposed by the benchmark, we evaluate on 300 (SWE-Bench Lite) and 500 (SWE-Bench Verified) samples to maintain consistency.

\subsection{RQ1: Efficacy of minimized regression tests.}
\label{sec:rq1}

\begin{figure}[t]
    \centering
    \includegraphics[width=0.87\columnwidth]{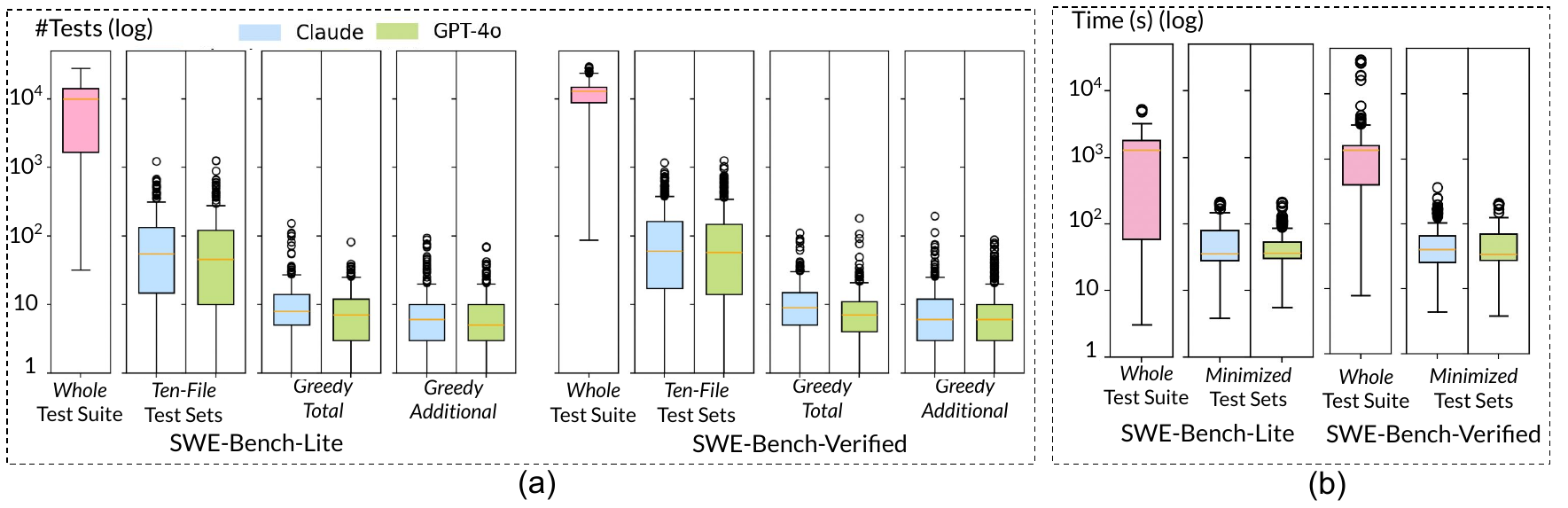}
    \vspace{-10pt}
    \caption{Comparison between the original test suite and \approach-minimized regression tests in terms of (a)~test size and (b) test runtime. The y-axis is shown on a logarithmic scale.
    }
    \label{fig:test-size}
    \vspace{-20pt}
\end{figure}

Figure~\ref{fig:test-size}-a shows the distribution of test sizes across three configurations: the complete test suites, the ten-file test sets, and the minimized regression tests generated by \approach.  The original repositories contain thousands of regression tests, with an average of $9{,}012$ tests for \lite and $11{,}769$ tests for \verified. Reducing to ten files significantly reduces the size of the test sets, yielding test suites that range from tens to hundreds of tests, with an overall average of $117$ across all models and datasets. Finally, with \approach, the number of executed tests is further reduced to an average of $9$ tests per instance under the greedy-additional algorithm and $11$ tests under the greedy-total algorithm. \approach reduces the size of tests by over a thousand times relative to the full test suites and by around $13$ times relative to the ten-file test sets. To demonstrate the effectiveness of reducing runtime cost, Figure~\ref{fig:test-size}-b reports the average runtime across all models and datasets. Running the full test suite takes 23m49s on average across all datasets and models. Among the projects, \texttt{scikit-learn} exhibits the highest average runtime at 37m31s. In contrast, executing our minimized regression tests requires only $52$ seconds on average, achieving a $27\times$ reduction in runtime compared to the original suite.

\vspace{.2cm}
\findings{1}{With only $9$ tests and an average runtime of $52$ seconds per instance, \approach achieves a reduction of over $1{,}000\times$ in test suite size and a $27\times$ improvement in runtime efficiency compared to the original full test suite.
}


\subsection{RQ2: Quality of localized test coverage.}
\label{subsec:rq2}

\begin{wrapfigure}{r}{0.55\columnwidth}
    \vspace{-10pt}
    \includegraphics[width=0.55\columnwidth]{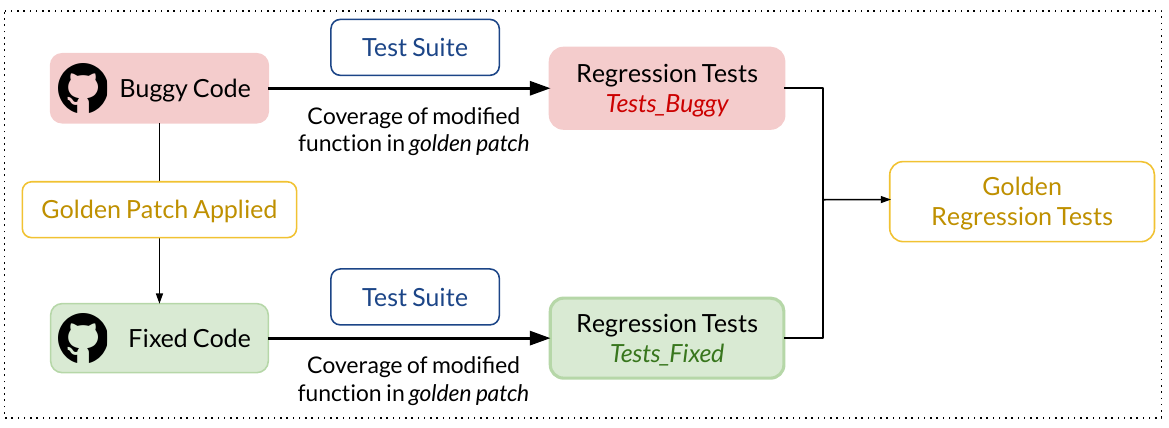}
    \vspace{-10pt}
    \caption{Constructing golden regression tests}
    \label{fig:golden-construct}
    \vspace{-10pt}
\end{wrapfigure}


To evaluate the effectiveness of our minimized regression tests, we construct a set of \textit{golden regression tests} as the baseline for test quality, defined as all tests that exercise the actual buggy functions. Figure~\ref{fig:golden-construct} illustrates this process: starting from the buggy version of the code in the dataset, we run the full test suite and collect line-level coverage information for all buggy functions. Every test that covers any buggy function is included in the regression test set \textit{tests\_buggy}. We then repeat the process on the fixed version of the code after applying the golden patch to obtain \textit{tests\_fixed}. The union of these two sets forms the final golden regression tests. As shown in Figure~\ref{fig:golden-stat}, the golden regression tests contain on average $322$ tests for \lite and $528$ tests for \verified.

We refer to the \approach minimized regression tests as \textit{MRT}, and the golden regression tests as \textit{GT}. We define two evaluation metrics of \textit{precision} and \textit{coverage recall} as follows:

\begin{equation}
\textit{Precision} = \frac{|\mathit{MRT} \cap \mathit{GT}|}{|\mathit{MRT}|}
\label{eq:precision}
\end{equation}

Let \( \mathcal{L}(T) \) denote the set of lines covered by a test set \( T \). Then, coverage recall is defined as:

\begin{equation}
\textit{Coverage Recall} = \frac{|\mathcal{L}(\mathit{MRT})|}{|\mathcal{L}(\mathit{GT})|}
\label{eq:recall}
\end{equation}

Precision measures how relevant the minimized regression tests are to the ground truth set of regression tests, while coverage recall quantifies the extent to which tests cover the buggy lines.

\begin{figure}[t]
    \centering
\includegraphics[width=0.8\columnwidth]{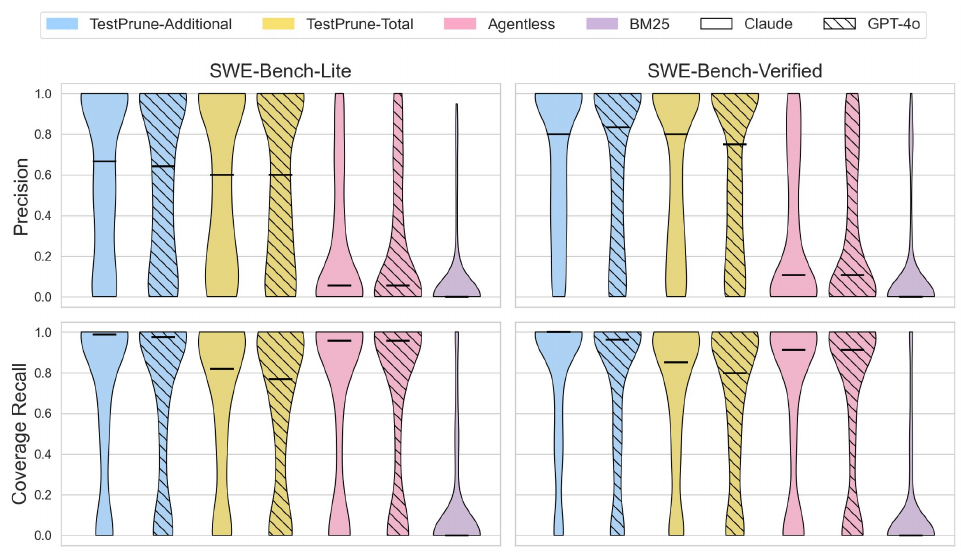}
\vspace{-10pt}
    \caption{Precision and coverage recall}
    \label{fig:intrinsic}
    \vspace{-20pt}
\end{figure}

We compare two variants of our approach, \approach-Additional and \approach-Total, against Agentless regression tests and a BM25-based baseline~\cite{bm25}. For the BM25 baseline, we retrieve the top $20$ most relevant tests using the issue description and the complete list of tests from the original repository. We choose a size of $20$ to match the typical size range of our minimized test sets. Figure~\ref{fig:intrinsic} shows the results: \approach-Additional achieves the highest precision, with an average of $0.63$, followed by \approach-Total at $0.60$. One possible reason for not reaching even higher precision is that our suspicious function sets may include more beyond the actual buggy function, thereby introducing more tests that may not exercise the actual buggy one.
Agentless has a lower precision of $0.28$, which can be attributed to its larger test set size, with an average of $79.5$ tests per instance. The BM25 baseline performs the worst, with an average precision of only $0.10$.

For coverage recall, \approach-Additional also outperforms all other methods, achieving an average of $0.71$ with a median of $0.98$ across all datasets and models. \approach-Total follows with an average coverage recall of $0.67$. Our greedy algorithm ensures that all lines covered by tests from the original ten selected test files are preserved, though these ten files may miss some lines compared to the full test suite. Agentless achieves a comparable coverage recall of $0.66$, while BM25 achieves only $0.13$. Despite having the smallest number of tests, \approach covers the largest number of relevant buggy lines. Since the greedy-additional algorithm demonstrates higher performance than greedy-total, we use the regression tests from \approach-Additional in the following experiments, and refer to \approach-Additional as \approach throughout.

\vspace{.2cm}
\findings{2}{\approach achieves the highest precision with an average of $0.63$ and the highest coverage recall of $0.71$, further confirming the effectiveness of \approach.}


\subsection{RQ3: Effectiveness of \approach on Issue Reproduction}
\label{result:rq3}

Table~\ref{tbl:otter} shows that our regression tests increase Otter’s fail-to-pass rate from 31.2\% to 34.0\% on SWE-Bench Verified with the Claude-3.7-Sonnet model. For other variants generated by masking localization information, the fail-to-pass rates range from 22.4\% to 27\%. Thus, Otter with \approach's minimized tests (``TestPrune'') achieves the best performance compared to the other variants.
Some recent works~\cite{wang_et_al_2025, nashid2025issue2test, ahmed2025execution, khatib2025assertflip} on reproduction test generation have achieved higher performance than Otter by leveraging execution feedback and inference scaling. In this work, however, our primary goal is to demonstrate the effectiveness of \approach for reproduction test generation. A direct comparison of TestPrune with other approaches is beyond the scope of this study. That said, we believe that other reproduction test generation techniques could also benefit from incorporating \approach.
We repeat the process with another model (GPT-4o) and on another benchmark (SWE-Bench Lite), and observe a similar performance improvement.

\approach improves the \mbox{$F\!\!\to\!\!P$ @ 1} rate, but for patch ranking  the more meaningful metric is \mbox{$F\!\!\to\!\!P$ @ N}, since six Otter-generated tests are applied in this setting.
In \( F\!\!\to\!\!P \) @ N, if at least one test among N changes from fail to pass, we count it as a success.
Table~\ref{tbl:otter-ablation} reports the  \mbox{$F\!\!\to\!\!P$ @ N} results for both models and benchmarks. On SWE-Bench Verified, Claude achieves a 49.8\% \mbox{$F\!\!\to\!\!P$ @ N} rate with N=6. This part of our work primarily focuses on how the \approach-supported variant, TestPrune, contributes to this \mbox{$F\!\!\to\!\!P$ @ N} performance. In addition to reporting the overall \mbox{$F\!\!\to\!\!P$ @ N} results in~\Cref{tbl:otter-ablation}, we also conduct an ablation study. In this experiment, we remove one variant at a time and measure \mbox{$F\!\!\to\!\!P$ @ 5}. We find that removing TestPrune has the largest negative impact. For example, on SWE-Bench Verified, performance drops by 8\%, indicating that TestPrune is the most influential variant. Moreover, TestPrune alone generates 20 unique fail-to-pass tests, whereas other variants generate only 3–10. We observe similar trends across other models and benchmarks.

\begin{wraptable}{r}{0.5\columnwidth}
\centering
\caption{Effectiveness of \approach on reproduction test generation}
\vspace{-12pt}
\resizebox{.50\columnwidth}{!}{%
\renewcommand{\arraystretch}{1.2}
\begin{tabular}{lllrr}
\toprule
\multicolumn{1}{c}{Benchmark}                                                  & \multicolumn{1}{c}{Model} & \multicolumn{1}{c}{Approach} & \multicolumn{1}{c}{\( F\!\!\to\!\!P \) @ 1 } & \multicolumn{1}{c}{Change} \\ \midrule
\multirow{4}{*}{\begin{tabular}[c]{@{}l@{}}SWE-Bench \\ Verified\end{tabular}} & \multirow{2}{*}{Claude}   & planner                        & 156(31.2\%)                     & NA                                                                            \\
                                                                               &                           & TestPrune              & 170(34.0\%)                     & +9.0\%                                                                             \\ \cline{2-5}
                                                                               & \multirow{2}{*}{GPT-4o}   & planner                        & 145(29.0\%)                      & NA                                                                            \\
                                                                               &                           & TestPrune               & 154(30.8\%)                     & +6.2\%                                                                           \\ \midrule
\multirow{4}{*}{\begin{tabular}[c]{@{}l@{}}SWE-Bench \\ Lite\end{tabular}}     & \multirow{2}{*}{Claude}   & planner                        & 81(27.0\%)                      & NA                                                                            \\
                                                                               &                           & TestPrune               & 88(29.3\%)                      & +8.6\%                                                                           \\ \cline{2-5}
                                                                               & \multirow{2}{*}{GPT-4o}   & planner                        & 71(23.7\%)                      & NA                                                                            \\
                                                                               &                           & TestPrune             & 76(25.3\%)                      & +7.0\%                                                                   \\  \bottomrule       
\end{tabular}
}
\label{tbl:otter}
\vspace{-10pt}
\end{wraptable}

\begin{table}[t]
\centering
\small
\caption{Impact of \approach on \( F\!\!\to\!\!P \) @ N}
\vspace{-8pt}
\resizebox{.85\columnwidth}{!}{%
\renewcommand{\arraystretch}{1.2}
\begin{tabular}{llr|rrr|rrr}
\toprule
\multicolumn{1}{c}{\multirow{2}{*}{Model}} & \multicolumn{1}{c}{\multirow{2}{*}{Prompt}} & \multicolumn{1}{c}{\multirow{2}{*}{N}} & \multicolumn{3}{c}{SWE-Bench Verified}                                                           & \multicolumn{3}{c}{SWE-Bench Lite}                                                               \\
\multicolumn{1}{c}{}                       & \multicolumn{1}{c}{}                        & \multicolumn{1}{c}{}                   & \multicolumn{1}{c}{\( F\!\!\to\!\!P \)} & \multicolumn{1}{c}{Unique Tests} & \multicolumn{1}{c}{Change} & \multicolumn{1}{c}{\( F\!\!\to\!\!P \)} & \multicolumn{1}{c}{Unique Tests} & \multicolumn{1}{c}{Change} \\ \midrule
\multirow{7}{*}{Claude}                    & All                                         & 6                                      & 249(49.8\%)                      & NA                               & NA                         & 131(43.7\%)                      & NA                               & NA                         \\
                                           & All - none                                  & 5                                      & 246(49.2\%)                      & 3                                & -1.2\%                       & 126(42.0\%)                        & 5                                & -3.8\%                       \\
                                           & All - testLoc                                & 5                                      & 244(48.8\%)                      & 5                                & -2.0\%                         & 126(42.0\%)                        & 5                                & -3.8\%                       \\
                                           & All - patchLoc                              & 5                                      & 241(48.2\%)                      & 8                                & -3.2\%                       & 124(41.3\%)                      & 7                                & -5.3\%                       \\
                                           & All - full                                  & 5                                      & 241(48.2\%)                      & 8                                & -3.2\%                       & 124(41.3\%)                      & 7                                & -5.3\%                       \\
                                           & All - planner                               & 5                                      & 239(47.8\%)                      & 10                               & -4.0\%                         & 127(42.3\%)                        & 4                                & -3.1\%                       \\
                                           & All - TestPrune                             & 5                                      & 229(45.8\%)                      & 20                               & -8.0\%                         & 123(41.0\%)                        & 8                                & -6.1\%                       \\ \midrule
\multirow{7}{*}{GPT-4o}                    & All                                         & 6                                      & 219(43.8\%)                      & NA                               & NA                         & 109(36.3\%)                      & NA                               & NA                         \\
                                           & All - none                                  & 5                                      & 215(43.0\%)                      & 4                                & -1.8\%                       & 106(35.3\%)                      & 3                                & -2.8\%                       \\
                                           & All - testLoc                                & 5                                      & 217(43.4\%)                      & 2                                & -0.9\%                       & 103(34.3\%)                      & 6                                & -5.5\%                       \\
                                           & All - patchLoc                              & 5                                      & 215(43.0\%)                      & 4                                & -1.8\%                       & 107(35.7\%)                      & 2                                & -1.8\%                       \\
                                           & All - full                                  & 5                                      & 215(43.0\%)                      & 4                                & -1.8\%                       & 106(35.3\%)                      & 3                                & -2.8\%                       \\
                                           & All - planner                               & 5                                      & 209(41.8\%)                      & 10                               & -4.6\%                       & 106(35.3\%)                      & 3                                & -2.8\%                       \\
                                           & All - TestPrune                             & 5                                      & 204(40.8\%)                      & 15                               & -6.8\%                       & 101(33.7\%)                      & 8                                & -7.3\%         \\ \bottomrule             
\end{tabular}
}
\label{tbl:otter-ablation}
\vspace{-20pt}
\end{table}

\vspace{2pt}

\noindent \textbf{How is \approach increasing the performance of Otter?} To reproduce an issue, the test must fail on the existing version of the code ($c_\mathrm{old}$). However, the test may not always fail for the reason described in the issue report. Chen et al.~\cite{chen2025unveiling} reported that Python execution errors during the issue resolution phase correlate with lower resolution rates. They also identified the most prevalent errors—such as ModuleNotFoundError and TypeError—in the generated code, which often differ significantly from the actual cause described in the issue. Providing a regression test that covers the suspicious function can help reduce such errors by supplying the necessary code syntax for those functions. It also gives the model a better understanding of relevant function usage and parameters.

Beyond syntax, \approach can mitigate the negative impact of incorrect test localization. Ahmed et al. reported that, for Otter, focal file localization accuracy is around 82\%, while test localization accuracy is about 70\%. The model performs better at identifying suspicious functions than at locating relevant tests. Incorporating regression tests can help the model compensate for errors in test localization. For example, in several instances (e.g., \textit{sympy\_\_sympy-20154, django\_\_django-13128, sphinx-doc\_\_sphinx-9281}), the model attempted to write a new test, but the test did not transition from fail to pass. In our experience, models are inherently better at modifying tests than at writing new ones from scratch. When we provided the \approach-selected tests, the model chose to modify an existing test instead of writing a new one, which enabled Otter to generate a proper fail-to-pass test. Thus, regression tests can play a crucial role in improving reproducibility.

While regression tests can help models by providing relevant context, they can also mislead models if the provided tests are irrelevant. When we consider our two primary variants—planner and TestPrune—their combined \mbox{$F\!\!\to\!\!P$ @ 2} is 205. Planner and TestPrune uniquely solve $35$ and $49$ fail-to-pass tests, respectively. This means TestPrune helps in 49 instances but also hurts performance in 35 cases. Nevertheless, \approach is beneficial overall. It also contributes by helping Otter generate  \( F\!\!\to\!\!P \) tests for additional instances for which none of the prior Otter variants could generate \( F\!\!\to\!\!P \) test. Note that the benefit in issue resolution can be increased by running multiple tests.

\vspace{.2cm}

\findings{3}{\approach increases the \( F\!\!\to\!\!P \) rate of Otter from 6.2\% to 9.0\%, regardless of the benchmark or model. The \approach-backed reproduction test generator TestPrune also increases the \( F\!\!\to\!\!P \) rate @ N by 6.0\% to 8.0\%.}
\vspace{-10pt}
\subsection{RQ4: Effectiveness of \approach on Issue Resolution}
\label{sec:rq4}

\begin{table}[t]
\centering
\caption{Impact of \approach on Issue Resolution}
\label{tbl:combined-resolution}
\vspace{-8pt}
\resizebox{.92\columnwidth}{!}{%
\renewcommand{\arraystretch}{1.18}
\begin{tabular}{lcccc|cc|rrr}
\toprule
\multicolumn{1}{c}{Benchmark} &
\multicolumn{1}{c}{Technique} &
\multicolumn{1}{c}{Model} &
\multicolumn{2}{c}{Regression Test} &
\multicolumn{2}{c}{Reproduction Test} &
\multicolumn{1}{c}{Resolved} &
\multicolumn{1}{c}{\% Resolved} &
\multicolumn{1}{c}{Change from Baseline} \\
& & & Default & \approach & Default & Otter & & & \\
\midrule
\multirow{10}{*}{\begin{tabular}[c]{@{}l@{}}SWE-Bench \\ Verified\end{tabular}}
& \multirow{4}{*}{Agentless}
& \multirow{4}{*}{Claude-sonnet-3.7}
& \cmark & \xmark & \cmark & \xmark & 254 & 50.8 & NA \\
& & & \xmark & \cmark & \cmark & \xmark & 260 & 52.0 & +2.4\% \\
& & & \xmark & \cmark & \xmark & \cmark & 268 & 53.6 & +5.5\% \\
& & & \xmark & \cmark & \cmark & \cmark & \textbf{278} & \textbf{55.6} & \textbf{+9.4\%} \\
\cline{2-10}
& \multirow{4}{*}{Agentless}
& \multirow{4}{*}{GPT-4o}
& \cmark & \xmark & \cmark & \xmark & 194 & 38.8 & NA \\
& & & \xmark & \cmark & \cmark & \xmark & 202 & 40.4 & +4.1\% \\
& & & \xmark & \cmark & \xmark & \cmark & 204 & 40.8 & +5.2\% \\
& & & \xmark & \cmark & \cmark & \cmark & \textbf{219} & \textbf{43.8} & \textbf{+12.9\%} \\

\cline{2-10}
& \multirow{2}{*}{Trae Agent}
& \multirow{2}{*}{Claude-sonnet-3.7}
& \cmark & \xmark & --- & --- & 325 & 65.0 & NA \\
& & & \xmark & \cmark & --- & --- & \textbf{351} & \textbf{70.2} & \textbf{+8.0\%} \\

\midrule
\multirow{10}{*}{\begin{tabular}[c]{@{}l@{}}SWE-Bench \\ Lite\end{tabular}}
& \multirow{4}{*}{Agentless}
& \multirow{4}{*}{Claude-sonnet-3.7}
& \cmark & \xmark & \cmark & \xmark & 122 & 40.7 & NA \\
& & & \xmark & \cmark & \cmark & \xmark & 127 & 42.3 & +4.1\% \\
& & & \xmark & \cmark & \xmark & \cmark & 126 & 42.0 & +3.3\% \\
& & & \xmark & \cmark & \cmark & \cmark & \textbf{135} & \textbf{45.0} & \textbf{+10.7\%} \\
\cline{2-10}
& \multirow{4}{*}{Agentless}
& \multirow{4}{*}{GPT-4o}
& \cmark & \xmark & \cmark & \xmark & 96 & 32.0 & NA \\
& & & \xmark & \cmark & \cmark & \xmark & 99 & 33.0 & +3.1\% \\
& & & \xmark & \cmark & \xmark & \cmark & 103 & 34.3 & +7.3\% \\
& & & \xmark & \cmark & \cmark & \cmark & \textbf{105} & \textbf{35.0} & \textbf{+9.4\%} \\

\cline{2-10}
& \multirow{2}{*}{SWE-agent}
& \multirow{2}{*}{Claude-sonnet-4}
& \cmark & \xmark & --- & --- & 170 & 56.7 & NA \\
& & & \xmark & \cmark & --- & --- & \textbf{186} & \textbf{62.0} & \textbf{+9.4\%} \\

\bottomrule
\end{tabular}
}
\vspace{-15pt}
\end{table}

\subsubsection{Effectiveness on non-agentic techniques.} Table~\ref{tbl:combined-resolution} \added{(rows \textit{Agentless})} shows the issue resolution rate of Agentless when using \approach-selected regression tests and Otter-generated reproduction tests. Note that Agentless uses both regression tests and reproduction tests in the patch ranking phase (discussed in~\Cref{background:agentless}). When we replace the default regression test set with \approach-selected tests, the issue resolution rate increases by 2.4\%–4.1\%. Although the GPT-4o model has a lower resolution rate than the Claude model, it benefits more from regression tests on SWE-Bench Verified. This suggests that running a small, relevant test set instead of a larger set with irrelevant tests can improve the issue resolution rate.
In~\Cref{result:rq3}, we have already shown that \approach improves the quality of the reproduction test generator, Otter. If we use Otter-generated tests instead of Agentless’s default tests with \approach, performance improves by 3.3\%–7.3\%. However, the best results are achieved when we use both Otter-generated and default reproduction tests. Applying both sets of tests requires minimal effort but results in a 9.4\%–12.9\% improvement in the issue resolution rate.

Unlike reproduction tests, regression tests cannot be combined. There is a high probability that \approach-selected tests are a subset of the default regression tests, so combining them would not add value. Furthermore, the primary goal of this paper is to obtain a minimized test set to reduce execution time and increase the issue resolution rate. Combining regression tests would diminish the benefits of \approach. 


We perform a McNemar test~\cite{mcnemar1947note} to show the effectiveness of \approach. It is a non-parametric statistical test used to compare paired proportions. It is applicable to SWE-Bench because the benchmark evaluates the correctness of two approaches on the same set of GitHub issues. This setup ensures their predictions are paired and directly comparable. We found statistical significance for all benchmark–model pairs at the 95\% confidence interval ($p < 0.05$). This significance holds for three benchmark–model pairs even at the 99\% level ($p < 0.01$). However, we did not observe statistical significance for SWE-Bench Lite using the GPT-4o model at 99\% confidence interval. This may be due to the fact that SWE-Bench Lite is a relatively small benchmark with only 300 samples. Overall, these findings confirm the effectiveness of our proposed approach to issue resolution.

\subsubsection{Effectiveness on agentic techniques.} 
\added{Table~\ref{tbl:combined-resolution} (rows \textit{Trae-agent} and \textit{SWE-agent}) also reports the results for agents. The reproduction test columns are marked as ``--'' because agents are only assisted by \approach regression tests and generate their own reproduction tests, without the help from Otter. Introducing \approach-selected minimized regression tests consistently improves issue resolution across both agents and benchmarks. On \verified, Trae-agent's resolved issues increase from 170 to 186 (+9.4\%), while on \lite, SWE-agent improves from 325 to 351 (+8.0\%). These improvements hold across different agents with different underlying models, suggesting that the benefit of minimized regression tests generalizes across diverse agents.}

\added{Beyond resolution rate, integrating \approach-selected regression tests also reduces both cost and the number of agent steps. SWE-agent exhibits a 23\% cost reduction (from \$1.18 to \$0.91 per instance on average) and 10\% fewer steps (from 64.6 to 58.3 steps) on \lite, while Trae-agent shows an 8\% cost reduction (from \$0.62 to \$0.57) and 5\% fewer steps (from 39.7 to 37.7 steps) on \verified. This can be due to the regression tests providing early and actionable feedback: when a candidate patch fails a regression test, the agent can immediately identify the problem and refine the fix, rather than pursuing a lot of exploratory steps that may lead to incorrect patches without the test run results. Regression tests serve as a tool that pushes the agent toward correct patches more efficiently.}
\added{The McNemar test also shows statistical significance for both agent–benchmark pairs at the 99\% confidence level ($p < 0.01$): SWE-agent on SWE-bench Lite ($p = 0.008$) and Trae-agent on SWE-bench Verified ($p = 0.001$), which shows that the improvement from integrating regression tests is unlikely due to the inherent non-determinism.}

\findings{4}{Replacing the default regression tests with the \approach-selected tests and incorporating the Otter reproduction tests can increase the issue resolution rate of Agentless by 9.4\%–12.9\% with statistical significance at 95\% confidence interval; \added{integrating \approach-selected tests further improves the issue resolution rate of Trae-agent and SWE-agent by 8.0\%–9.4\%, regardless of the benchmark and model. }}

\vspace{-10pt}
\section{Discussion}
\label{sec:discussion}

\subsection{Cost Effectiveness of \approach}

Agentless makes several calls to LLMs for localization, reproduction test generation, regression test filtering, and candidate generation. The authors reported a cost of \$0.70 per sample for the GPT-4o model~\cite{xia_et_al_2025}. The cost can go up to \$1.50 with models like Claude-Sonnet-3.7. Another client of \approach, Otter, costs only \$0.09 for each instance with the GPT-4o model~\cite{ahmed_et_al_2025}; with the additional cost for the \approach variant, the total cost is \$0.11 for each instance. Therefore, with the GPT-4o model, the total cost for Agentless plus Otter is \$0.81.
In \approach, we primarily make one LLM call to retrieve test files and a few additional calls for tie-breaking tests with the same coverage. From our estimation, for GPT-4o, this should cost \$9.50 for the 500 instances of SWE-Bench Verified, which results in \$0.02 per instance. Thus, the total cost for \approach and the downstream tasks per instance is less than \$0.85 with GPT-4o, showing that \approach hardly introduces any additional cost. For Claude, \approach also costs less than \$0.05 per instance.

\vspace{-10pt}
\subsection{Coverage of Reproduction Tests on SWE-Bench Verified}

\begin{wrapfigure}{r}{0.55\columnwidth}
    \vspace{-10pt}
\includegraphics[width=0.55\columnwidth]{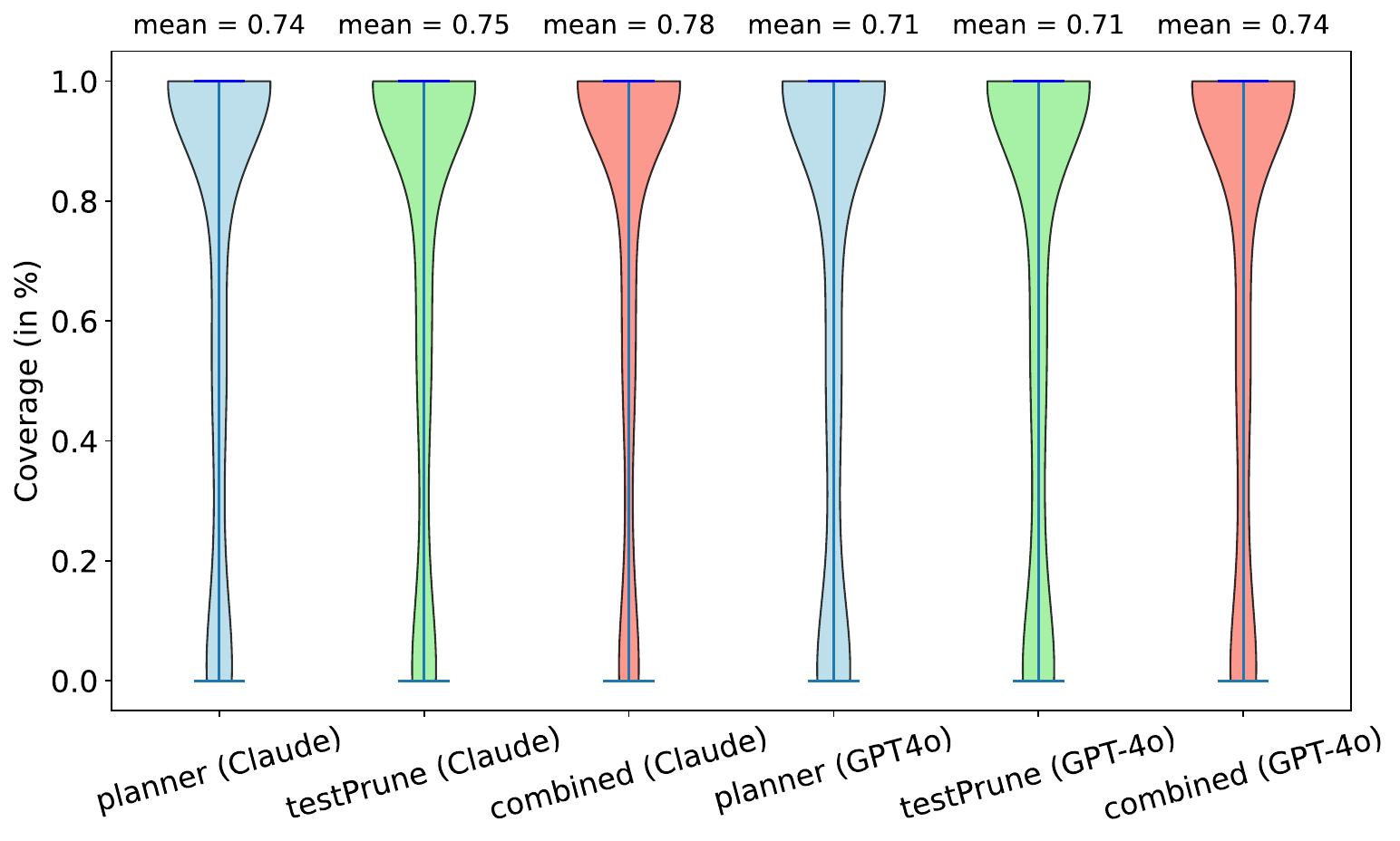}
\vspace{-20pt}
    \caption{Coverage of the reproduction test}
    \vspace{-15pt}
    \label{fig:coverage}
\end{wrapfigure}

Coverage is used to measure the adequacy of tests. We measure how well a generated test covers the developer-written golden code patch (note that this is not available during test generation) for planner and TestPrune variants. To compute coverage, we locate the deleted lines in $c_\mathrm{old}$ and the added lines in $c_\mathrm{new}$ and see whether the test covers those lines. We take the total number of covered lines and divide them by the total number of deleted and added lines.
Note that although regression tests are selected based on the coverage of the suspicious function(s), they are unlikely to increase the coverage of the reproduction test. The reproduction test only uses them in the planning phase and as additional context. Though the  \mbox{$F\!\!\to\!\!P$} rate increases for all benchmark and model pairs, the coverage of the test does not increase much. Figure~\ref{fig:coverage} shows that on SWE-Bench Verified with the Claude model, the coverage increased by a negligible amount for TestPrune.
Prior research shows that the coverage of a test is directly linked to the success of the test~\cite{mundler_et_al_2024, ahmed_et_al_2025}. We may be able to use these minimized regression tests to increase the coverage of the reproduction test, which will increase the quality of the reproduction test. We leave this for future research.
These findings on coverage are somewhat consistent with our observations on the reproduction test set. The TestPrune variant generates only 14 more \mbox{$F\!\!\to\!\!P$} tests than planner with Claude on SWE-Bench Verified, so the coverage is unlikely to increase. However, if we apply both tests (discussed in~\Cref{result:rq3}), both the  \mbox{$F\!\!\to\!\!P$} rate ($156 \rightarrow 205$) and coverage go up ($0.74 \rightarrow 0.78$). To compute combined coverage, we take the coverage of planner and TestPrune and keep the higher one.

\vspace{-10pt}
\subsection{Analysis of Issue Resolution}

\subsubsection{Analysis of Agentless Patch Selection.} 
\begin{wrapfigure}{r}{0.42\columnwidth}
    \vspace{-20pt}
\includegraphics[width=0.42\columnwidth]{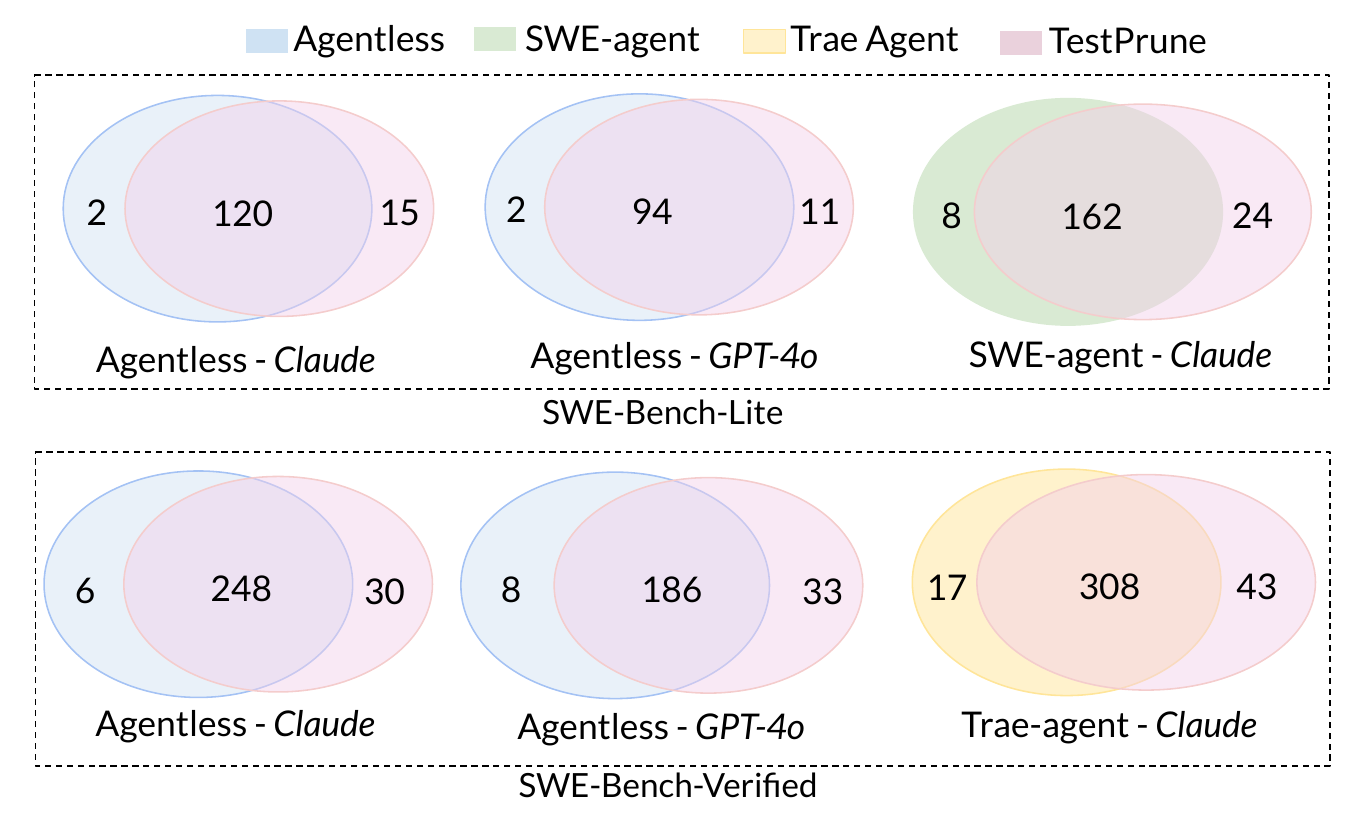}
\vspace{-24pt}
    \caption{Comparison between resolved instances of issue resolution with \approach}
    \vspace{-20pt}
    \label{fig:venn}
\end{wrapfigure}

To further investigate the patch selection results, Figure~\ref{fig:venn} shows the overlap between the correct patches selected by \approach and those by Agentless. On the SWE-Bench Lite dataset, \approach exclusively fixes $15$ instances with Claude and $11$ with GPT-4o, while on the SWE-Bench Verified dataset, it exclusively fixes $30$ and $33$ instances, respectively.

\approach regression tests assist patch selection by running more relevant tests; patches that fail to preserve correct functionality can be excluded (e.g., those failing on newly added tests). For example, for the instance \texttt{\footnotesize django\_\allowbreak\_django-11964} in SWE-Bench Lite, \approach was able to validate the correct patch while Agentless was not. In this case, the \approach test set included five tests from classes absent in the Agentless test set (e.g., \texttt{\footnotesize  \footnotesize{model\_fields\allowbreak.test\_integerfield\allowbreak.IntegerFieldTests}}),
which successfully failed the incorrect patch chosen by Agentless and allowed the correct patch for a higher rank. 

Otter further helps patch selection by providing additional reproduction tests, which increase the likelihood of identifying correct patches when the original reproduction test fails. For example, in \texttt{\footnotesize  \footnotesize django\_\_django-14608}, Agentless was unable to reproduce the issue, whereas a reproduction test generated by Otter successfully reproduced it, enabling the correct patch to be returned. Including multiple reproduction tests also helps prevent patch selection from depending too much on a single reproduction test, which might be wrong, and instead increases the likelihood of returning a good patch.

We fail to validate some instances that the original Agentless can handle, largely for the same reasons as other failures: many of these cases are caused by tie-breaking issues, where multiple patches achieve the same test passing rate; in such cases, majority voting favors patches with higher occurrences. In addition, some instances are misled by reproduction tests that prioritize the wrong patch. Nevertheless, as shown in ~\Cref{sec:rq4}, \approach benefits the issue resolution in general.

\vspace{-10pt}
\subsubsection{Analysis of Agent Issue Resolution} 
\added{Figure~\ref{fig:venn} shows the comparison of resolved instances before and after applying \approach. On \lite, SWE-agent exclusively resolves 24 new instances, while on \verified, Trae Agent exclusively resolves 43 new instances.
We already saw an example in Figure~\ref{fig:test-design}~(b).
In addition, we identify two other main ways in which regression tests improve agent performance.}

\added{\approach can help agents localize bugs and write better reproduction tests. 
For example, in instance \texttt{\footnotesize django-11422}, the issue was that Django's autoreloader skipped \texttt{\footnotesize manage.py} because it runs as \texttt{\footnotesize \_\_main\_\_}, which lacks the \texttt{\footnotesize \_\_spec\_\_} attribute. 
Without regression tests, the original SWE-agent mislocalized the bug, it patched a different function \texttt{\footnotesize run\_with\_reloader} which never resolved the actual issue, and spent 28~steps writing ineffective test scripts, ultimately failing to reproduce or fix the issue. 
With regression tests integrated,
the agent runs related regression tests early (\texttt{\footnotesize  TestIterModulesAndFiles.test\_file\_added}), which pointed the agent toward \texttt{\footnotesize  iter\_modules\_and\_files} as the relevant function. Reading the test code revealed how existing tests called functions (\texttt{\footnotesize iter\_all\_python\_module\_files}) with path resolution to assert file tracking, guiding the agent to write a reproduction script following the same pattern. This led the agent to pinpoint the \texttt{\footnotesize \_\_spec\_\_ is None} guard as the root cause and produce a correct fix. The regression tests served a triple role: their names helped localize the bug, their code patterns guided reproduction test construction, and their assertions validated the fix.
}

\added{\approach can also help agents converge faster and avoid spurious costs. For example, in instance \texttt{\footnotesize django-16408}, the issue was that multi-level \texttt{\footnotesize FilteredRelation} paths combined with \texttt{\footnotesize select\_related()} caused Django to cache the wrong related object. Both agents initially applied the same incorrect fix: disabling \texttt{\footnotesize local\_setter} for all \texttt{\footnotesize FilteredRelation}, which fixed multi-level cases but broke single-level ones.
With regression tests, the agent had already seen relevant tests (e.g., \texttt{\footnotesize test\_select\_related}) at the beginning. After its first patch, it ran these tests and immediately discovered the single-level cases broken at step~51. The failing test clarified that the fix should depend on whether the \texttt{\footnotesize FilteredRelation} definition itself spans multiple hops, leading to a correct fix by step~57.
Without regression tests, the original agent spent 63~steps running the only test mentioned in the issue description and writing some debugging scripts from scratch. It did not discover the single-level breakage until step~64, triggering repeated fix-break-revert cycles over 50 more additional steps, and ultimately produced an incorrect fix. Overall, integrating \approach regression tests reduced this instance from 120~steps (\$3.74) to 84~steps (\$1.94) while changing the outcome from failure to success.}

\added{Integrating minimized regression tests also caused some previously resolved instances to fail---8 for SWE-agent and 17 for Trae Agent. The main root cause is that the selected regression tests covered general functionality but missed the specific buggy behavior. When all tests passed, the agent treated this as validation success and stopped without further exploration. Without regression tests, the agent would have continued exploring and potentially found the correct fix.}
\vspace{-10pt}
\section{Threats To Validity}
\label{sec:threats}






One of the major limitations of this work is that our experiments are limited to Python only. SWE-Bench includes 12 repositories, so our findings may not generalize to other repositories or programming languages. However, works like SWE-Bench~\cite{jimenez_et_al_2024}, Agentless~\cite{xia_et_al_2025}, AutoCodeRover~\cite{zhang2024autocoderover}, and CodeMonkeys~\cite{ehrlich2025codemonkeys} also have this limitation, yet they have still been impactful and contributed significantly to the field.

Model contamination/model memorization is a serious concern in SWE-Bench related work. The model has likely already seen the code repository during pretraining and may thus perform well on these samples. SWE-Bench~\cite{jimenez_et_al_2024}, SWT-bench~\cite{mundler_et_al_2024}, and Otter~\cite{ahmed_et_al_2025}  have addressed this problem and argued that the impact of contamination is limited. The SWE-Bench  paper claims that ``difficulty does not correlate with issue resolution date''.  Ahmed et al.~\cite{ahmed_et_al_2025} show that Otter-generated tests  differ substantially from developer-generated tests.
In this work, we use Otter-generated tests and Agentless-generated code patches. We are not introducing any additional contamination in the process. In \approach, we primarily make one LLM call to retrieve the 10 most relevant test files, which is a relatively easier task and works based on issue and file-name matching. It does not require complex reasoning or syntactical correctness, unlike test and code patches. After selecting the 10 files, the rest of the algorithm makes no further LLM calls. Since \approach makes one LLM call, which does not generate code, we believe our results are not highly impacted by the model contamination problem. We also make some additional calls for tie-breaking tests with the same coverage. These calls are very similar to test file localization and do not write any functions.

\approach uses a greedy algorithm on the output of the Python ``coverage'' package. Greedy algorithms are simple to run, but they certainly have some limitations and are not capable of providing an optimal solution. However, since we achieve high precision and coverage, we can infer that our results are not significantly impacted by the lack of an optimal solution.
Additionally, the ``coverage'' package sometimes fails to generate correct coverage reports due to dependency issues or parallel execution. Since we use a pre-built Docker container, we did not encounter many dependency issues, and coverage failed for fewer than 1\% of samples.

Finally, LLMs sometimes generate different solutions even at temperature 0. Agentless and Otter have reported their numbers on different benchmarks using different models and found that their approaches are effective in general. In our approach for test file localization, we tried several runs but did not see noticeable changes in the output. Note that the order of test files does not matter in our setting. We use the coverage package, which neutralizes the impact of file ordering. Additionally, the test file detection @ 10 accuracy is very high in our case (around 95\%). From these observations, we believe that our results will not change much even if we re-run our experiments. Besides, we tried multiple models and benchmarks to show that our approach is effective in general.

\vspace{-10pt}
\section{Related Work}
\label{sec:relatedwork}

This paper is the first to formulate issue-based test minimization and to provide a dedicated solution. Prior work has addressed related tasks that involve issues, often leveraging existing regression tests along the way.
Agentic software engineering techniques take an issue description and a repository as input, and output a patch candidate. 
Agentless~\cite{xia_et_al_2025} selects regression tests (old tests) in two steps: filtering to keep only tests that pass on buggy code, then applying an LLM filter. It also uses the selected tests to rank candidate patches. SpecRover~\cite{ruan_zhang_roychoudhury_2025} executes regression tests on each patch and uses an LLM to interpret results. SWE-RL~\cite{wei_et_al_2025}, R2E-Gym~\cite{jain_et_al_2025}, PatchPilot~\cite{li_et_al_2025}, and Trae Agent~\cite{gao2025trae} all adopt regression testing similar to Agentless, differing mainly in how selected tests are applied to filter or rank candidate patches. Unlike these approaches, \approach incorporates coverage into test selection, applies systematic minimization, and evaluates selection not only for issue reproduction but also through intrinsic metrics. Our empirical results highlight the limitations of prior work, improving  Agentless end-to-end as a proof of concept. 

A second line of work focuses on generating new tests to reproduce issues. While distinct from selecting old tests, several systems combine both. Otter++\cite{ahmed_et_al_2025} selects old tests by prompting an LLM at both file and function granularity, then includes them in prompts for generating new tests. AEGIS~\cite{wang_et_al_2025} uses a ReAct agent to locate issue-related code, including tests, and places it in a prompt context for test generation. USEAgent~\cite{applis_et_al_2026} employs a meta-agent that can invoke sub-agents for test retrieval and execution, using results to guide patch refinement. Similarly, BRT Agent~\cite{cheng_et_al_2025} retrieves old tests via code search and includes them in new test generation prompts. Unlike these systems, \approach uniquely leverages coverage for test selection, applies minimization, and evaluates test selection in isolation. We demonstrate that \approach improves Otter++. 

There is also related research on localizing focal functions (not tests) from issues—a subtask that strongly benefits SWE agents. OrcaLoca~\cite{yu_et_al_2025} uses search APIs for focal localization. LocAgent~\cite{chen_et_al_2025} builds a code index and applies an agent with a graph traversal API, while CoSIL~\cite{jiang_et_al_2025} constructs the graph index just-in-time for efficiency. SweRank~\cite{reddy_et_al_2025} fine-tunes two models: one for embedding-based retrieval and another for ranking focal candidates. In contrast to these function localizers, \approach focuses on localizing tests, exploiting test-specific properties such as coverage.

Test suite minimization has also been widely used to reduce the cost of regression testing while preserving original test suite capabilities~\cite{yoo2012regression}. Existing works have explored the use of greedy algorithms~\cite{jabbarvand_et_al_2016,hsu2009mints,jones2003test,xie2004checking,jeffrey2005test}, integer programming~\cite{lin2018nemo,jabbarvand_et_al_2016}, meta-heuristic search~\cite{black2004bi,gu2025scalable,wang2015cost,pan2023atm}, or machine learning~\cite{vahabzadeh2018fine,pan2023atm,pan2024ltm} to select the minimal subset of test suites with comparable effectiveness. Greedy algorithms are fast, but provide a near-optimal solution. The integer programming and meta-heuristic search can take a longer time to terminate, but provide the optimal solution. Prior works mostly compare the effectiveness of the original and minimized test suites concerning code coverage (lines, branches, and predicates), requirement coverage, execution time, and fault-detection abilities. Concerning preserving the fault-detection abilities of test suites, they rely on historical data or runtime execution under mutation testing. To the best of our knowledge, none of the prior work has explored or implemented test suite minimization concerning \emph{debugging} abilities of \emph{future} faults. This criterion is also adaptive, i.e., test suite minimization should be performed per issue, requiring a scalable yet effective approach for quick test suite minimization. 

\vspace{-12pt}
\section{Conclusion}
\label{sec:conclusion}

This paper makes three contributions:
(a)~formulating the problem of issue-driven test minimization,
(b)~introducing \approach, a workflow that solves this problem; and
(c)~showing how to leverage the resulting test subset
for issue reproduction and issue resolution.
\approach aims to minimize test subset size
while maximizing relevance as per our new coverage recall metric.
While some prior approaches include a basic test selection step as
part of a larger workflow, they do not leverage coverage, nor do they
measure, let alone optimize, the intrinsic performance of this step.
\added{This paper shows that using regression tests selected by
\approach improves the performance of diverse downstream
clients~(Otter++, Agentless, SWE-agent, and Trae Agent).}
Overall, this work complements prior work on test minimization by
showing how to cover the new code that resolves an issue without
access to that code. The artifacts of \approach are publicly available at ~\cite{artifact}.




\bibliographystyle{ACM-Reference-Format}
\bibliography{ref}

\end{document}